\documentclass[apj]{emulateapj}
\usepackage{apjfonts}
\usepackage{lscape}

\shorttitle{The Gemini Deep Planet Survey}
\shortauthors{Lafreni\`{e}re et al.}

\newcommand{\mjup}{\ensuremath{M_{\rm Jup}}}
\newcommand{\drm}{\ensuremath{\mathrm{d}}}
\newcommand{\mmin}{\ensuremath{m_{\rm min}}}
\newcommand{\mmax}{\ensuremath{m_{\rm max}}}
\newcommand{\amin}{\ensuremath{a_{\rm min}}}
\newcommand{\amax}{\ensuremath{a_{\rm max}}}
\newcommand{\lkhd}{\ensuremath{\mathcal{L}}}
\newcommand{\fmax}{\ensuremath{f_{\rm max}}}
\newcommand{\fmin}{\ensuremath{f_{\rm min}}}

\begin{document}

\title{The Gemini Deep Planet Survey -- GDPS\footnote{\lowercase{\uppercase{B}ased on observations obtained at the \uppercase{G}emini \uppercase{O}bservatory, which is operated by the \uppercase{A}ssociation of \uppercase{U}niversities for \uppercase{R}esearch in \uppercase{A}stronomy, \uppercase{I}nc., under a cooperative agreement with the \uppercase{NSF} on behalf of the \uppercase{G}emini partnership: the \uppercase{N}ational \uppercase{S}cience \uppercase{F}oundation (\uppercase{U}nited \uppercase{S}tates), the \uppercase{P}article \uppercase{P}hysics and \uppercase{A}stronomy \uppercase{R}esearch \uppercase{C}ouncil (\uppercase{U}nited \uppercase{K}ingdom), the \uppercase{N}ational \uppercase{R}esearch \uppercase{C}ouncil (\uppercase{C}anada), \uppercase{CONICYT} (\uppercase{C}hile), the \uppercase{A}ustralian \uppercase{R}esearch \uppercase{C}ouncil (\uppercase{A}ustralia), \uppercase{CNP}q (\uppercase{B}razil) and \uppercase{CONICET} (\uppercase{A}rgentina).}}}

\author{
David Lafreni\`ere\altaffilmark{a},
Ren\'e Doyon\altaffilmark{a},
Christian Marois\altaffilmark{b},
Daniel Nadeau\altaffilmark{a},
Ben R. Oppenheimer\altaffilmark{c},
Patrick F. Roche\altaffilmark{d},
Fran\c cois Rigaut\altaffilmark{e},
James R. Graham\altaffilmark{f},
Ray Jayawardhana\altaffilmark{g},
Doug Johnstone\altaffilmark{h},
Paul G. Kalas\altaffilmark{f},
Bruce Macintosh\altaffilmark{b},
Ren\'e Racine\altaffilmark{a}
}
\altaffiltext{A}{D\'epartement de physique and Observatoire du Mont M\'egantic, Universit\'e de Montr\'eal, C.P. 6128 Succ. Centre-Ville, Montr\'eal, QC, H3C 3J7, Canada}
\altaffiltext{B}{Institute of Geophysics and Planetary Physics L-413, Lawrence Livermore National Laboratory, 7000 East Ave, Livermore, CA 94550}
\altaffiltext{C}{Department of Astrophysics, American Museum of Natural History, Central Park West at 79th Street, New York, NY 10024}
\altaffiltext{D}{Astrophysics, Physics Department, University of Oxford, 1 Keble Road, Oxford, OX1 3RH, UK}
\altaffiltext{E}{Gemini Observatory, Southern Operations Center, Association of Universities for Research in Astronomy, Inc., Casilla 603, La Serena, Chile}
\altaffiltext{F}{Department of Astronomy, University of California at Berkeley, 601 Campbell Hall, Berkeley, CA 94720}
\altaffiltext{G}{Department of Astronomy and Astrophysics, University of Toronto, 50 St. George Street, Toronto, ON, M5S 3H4, Canada}
\altaffiltext{H}{National Research Council Canada, Herzberg Institute of Astrophysics, 5071 West Saanich Road, Victoria, BC, V9E 2E7, Canada}
\email{david@astro.umontreal.ca doyon@astro.umontreal.ca}



\begin{abstract} 

We present the results of the Gemini Deep Planet Survey, a near-infrared adaptive optics search for giant planets and brown dwarfs around nearby young stars. The observations were obtained with the Altair adaptive optics system at the Gemini North telescope and angular differential imaging was used to suppress the speckle noise of the central star. Detection limits for the 85 stars observed are presented, along with a list of all faint point sources detected around them. Typically, the observations are sensitive to angular separations beyond 0.5\arcsec\ with 5$\sigma$ contrast sensitivities in magnitude difference at 1.6~$\mu$m of 9.5 at 0.5\arcsec, 12.9 at 1\arcsec, 15.0 at 2\arcsec, and 16.5 at 5\arcsec. For the typical target of the survey, a 100~Myr old K0 star located 22~pc from the Sun, the observations are sensitive enough to detect planets more massive than 2~\mjup\ with a projected separation in the range 40--200~AU. Depending on the age, spectral type, and distance of the target stars, the detection limit can be as low as $\sim$1~\mjup. Second epoch observations of 48 stars with candidates (out of 54) have confirmed that all candidates are unrelated background stars. A detailed statistical analysis of the survey results, yielding upper limits on the fractions of stars with giant planet or low mass brown dwarf companions, is presented. Assuming a planet mass distribution $\drm n/\drm m \propto m^{-1.2}$ and a semi-major axis distribution $\drm n/\drm a \propto a^{-1}$, the 95\% credible upper limits on the fraction of stars with at least one planet of mass 0.5--13~\mjup\ are 0.28 for the range 10--25~AU, 0.13 for 25--50~AU, and 0.093 for 50--250~AU; this result is weakly dependent on the semi-major axis distribution power-law index. The 95\% credible interval for the fraction of stars with at least one brown dwarf companion having a semi-major axis in the range 25--250~AU is $0.019_{-0.015}^{+0.083}$, irrespective of any assumption on the mass and semi-major axis distributions. The observations made as part of this survey have resolved the stars HD~14802, HD~166181, and HD~213845 into binaries for the first time.

\end{abstract} 
\keywords{Planetary systems --- stars: imaging --- binaries: close --- stars: low-mass, brown dwarfs}

\section{Introduction}

\setcounter{footnote}{0}

More than 200 exoplanets have been discovered over the last decade through precise measurements of variations of the radial velocity (RV) of their primary star. Besides establishing that at least 6--7\% of FGK stars have at least one giant planet with a semi-major axis smaller than $\sim$5~AU \citep{marcy05}, the profusion of data following from the RV discoveries has propelled the field of giant planet formation and evolution into an unprecedented state of activity. For a review of the main characteristics of the RV exoplanets, the reader is referred to \citet{udry07,butler06,marcy05}. Besides the RV technique, the photometric transit method has lead successfully to the discovery of new exoplanets on small orbits \citep[e.g.][]{konacki03,alonso04,cameron07} and has provided the first measurements of the radius and mean density of giant exoplanets \citep[e.g.][]{charbonneau00}. Very recently, a few exoplanets have been detected by gravitational microlensing \citep{bond04,udalski05,beaulieu06, gould06}; these planets have separations of $\sim$2--5~AU. Notwithstanding their great success in finding planets on small orbits, these techniques cannot be used to search for and characterize planets on orbits larger than $\sim$10~AU. As a result, the population of exoplanets on large orbits is currently unconstrained.

The two main models of giant planet formation are core accretion \citep{pollack96} and gravitational instability \citep{boss97,boss01}. In the core accretion model, solid particles within a proto-planetary disk collide and grow into solid cores which, if they become massive enough before the gas disk dissipates, trigger runaway gas accretion and become giant planets. Models predict that the timescale for formation of a planet like Jupiter through this process is about 5~Myr \citep{pollack96}, or about 1~Myr if migration of the core through the disk is allowed as the planet forms \citep{alibert05}. These timescales are comparable to or below the estimated proto-planetary dust disk lifetime ($\sim$6~Myr, \citealp{haisch01}) and gas disk lifetime ($\lesssim$10~Myr, \citealp{jayawardhana06}). Formation through core accretion is strongly dependent on the surface density of solid material (hence [Fe/H])  in the proto-planetary disk , precluding formation of Jupiter mass planets at distances greater than 15--20~AU \citep[e.g.][]{pollack96, ida04}, where the low density of planetesimals would lead to prohibitively long formation timescales. Neptune mass planets can be formed out to slightly larger distances and can further migrate outward owing to interaction with the disk.

In the gravitational instability model, small instabilities in a proto-planetary disk grow rapidly into regions of higher density that subsequently evolve into spiral arms owing to Keplerian rotation. Further interactions between these spiral arms lead to the formation of hot spots which then collapse to form giant planets. The range of orbital separation over which this mechanism may operate efficiently is not yet clear. Some studies indicate that it may lead to planet formation only at separations exceeding $\sim$100~AU \citep{whitworth06, matzner05}, where the radiative cooling timescale is sufficiently short compared to the dynamical timescale, while others have been able to produce planets only at separations below 20--30~AU \citep{boss00, boss03, boss06}.

A few other models are capable of forming giant planets on large orbits directly. One such mechanism is shock-induced formation following collision between disks \citep{shen06}. In this model, the violent collision of two proto-planetary disks triggers instabilities that lead to the collapse of planetary or brown dwarf (BD) mass clumps. Results of numerical simulations indicate that planets and BDs may form at separations of several tens of AU or more through this process \citep{shen06}. The competitive accretion and ejection mechanism that was proposed initially to explain the formation of BDs \citep{reipurth01} could also form planetary mass companions on large orbits, as suggested by the results of recent simulations by \citet{bate05}.

Even in a scenario in which all giant planets form on small orbits, through either core accretion or gravitational collapse, a significant fraction of planets could be found on stable orbits of tens of AU because of outward orbital migration. Indeed, numerical simulations have shown that gravitational interactions between planets in a multi-planet system may send one of the planets, usually the least massive one, out to an eccentric orbit of semi-major axis of tens to hundreds of AU \citep{chatterjee07, veras04, rasio96, weidenschilling96}. This process could be involved frequently in the shaping of the orbital parameters of planetary systems as we have learned from RV surveys that multi-planet systems are common, representing $\sim$14\% of known planetary systems \citep{marcy05}. Outward migration of massive planets can be induced also by interactions between the planet and the gaseous disk; the simulations of \citet{veras04} reveal that this process is capable of carrying Jupiter mass planets out to several tens of AU. Similarly, angular momentum exchange between two planets (or more), achieved through viscous interactions with the disk, could drive the outer planet to a separation of hundreds of AU \citep{martin07}. Outward planet migration can result further from interaction of the planet with the solid particles in the disk after the gas has dissipated \citep[e.g.][]{levison07}; there is in fact strong evidence that this mechanism has played an important role in the Solar system \citep{fernandez84,malhotra95,hahn05}. Based on numerical simulations, it is likely that all giant planets of the Solar system formed interior to $\sim$15~AU and migrated outward (except Jupiter) to their current location \citep{tsiganis05}.

From an observational point of view, there is some evidence that planets on large orbits may exist. Many observations of dusty disks around young stars, made either in emitted light (e.g. Vega, $\varepsilon$~Eri, Fomalhaut; \citealp{holland98, greaves98}) or in scattered light (e.g. HD~141569, HR~4796, Fomalhaut; \citealp{augereau99, weinberger99, schneider99, kalas05}), have unveiled asymmetric or ring-like dust distributions. These peculiar morphologies could arise from gravitational dust confinement imposed by one or more (unseen) giant planets on orbits of tens to hundreds of AU. In fact, detailed numerical simulations of the effect of giant planets on the dynamical evolution of dusty disks have been able to reproduce the observed morphologies with remarkable agreement \citep{ozernoy00, wilner02, deller05}. Typically, Jupiter mass planets on orbits of $\sim$60~AU are needed to reproduce the observations, although in some cases less massive planets (similar to Neptune) may be able to reproduce the observed features.

In the last few years, there have been a few discoveries of planetary mass or low-mass BD companions located beyond several tens of AU, in projection, from their primary: an $\sim$8~\mjup\ companion 40~AU from the BD 2M~1207$-$3932 \citep{mohanty07,chauvin05a,chauvin04}, a $\sim$25~\mjup\ companion 100~AU from the T~Tauri star GQ~Lup \citep{marois07, seifahrt07, neuhauser05}, a $\sim$12~\mjup\ companion 210~AU from the young star CHXR~73 \citep{luhman06}, a $\sim$25~\mjup\ companion 240~AU from the young BD 2M~1101$-$7732 \citep{luhman04}, a $\sim$12~\mjup\ companion 260~AU from the young star AB~Pic \citep{mohanty07, chauvin05b}, a 7--19~\mjup\ companion 240--300~AU from the young BD Oph~1622$-$2405 \citep{luhman07b,close07,jayawardhana06}, an $\sim$11~\mjup\ companion 330~AU from the T~Tauri star DH~Tau \citep{luhman06, itoh05}, and a $\sim$21~\mjup\ companion 790~AU from the star HN~Peg \citep{luhman07}. These discoveries might indicate that more similar companions, and less massive ones, do exist and remain to be found.

Perhaps even more compelling is the fact that the number of exoplanets found by RV surveys increases as a function of semi-major axis for the range 0.1--3~AU \citep{butler06}; these surveys are incomplete at larger separations. Conservative extrapolation suggests that there may be at least as many planets beyond 3~AU as there are within \citep{butler06}. In fact, long-term trends in RV data have been detected for about 5\% of the stars surveyed \citep{marcy05}, suggesting the presence of planets between 5~AU and 20~AU around them.

\begin{figure*}[t]
\epsscale{1}
\plotone{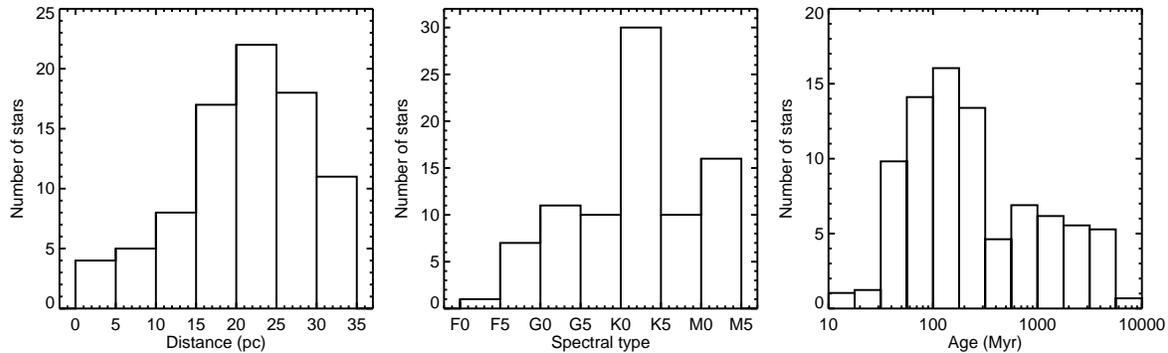}
\caption{\label{fig:targets} Distribution of distance, spectral type, and age of the target stars. For the age distribution, each star was distributed over all the age bins according to the fraction of their estimated age interval falling inside each bin.}
\end{figure*}

Given all of the considerations above, it is clear that a determination of the frequency of giant planets as a function of orbital separation out to hundreds of AU is necessary to elucidate the relative importance of the various modes of planet formation and migration. Direct imaging is currently the only viable technique to probe for planets on large separations and achieve this goal. However, detecting giant planets directly through imaging is very difficult due to the angular proximity of the star and the very large luminosity ratios involved. Currently, the main technical difficulty when trying to image giant planets directly does not come from diffraction of light by the telescope aperture, from light scattering due to residual atmospheric wavefront errors after adaptive optics (AO) correction, nor from photon noise of the stellar point spread function (PSF), but rather from light scattering by optical imperfections of the telescope and camera that produce bright quasi-static speckles in the PSF of the central star. These speckles are usually much brighter than the planets sought after. More in depth discussions of this problem, as well as possible venues to circumvent it using current instrumentation, can be found in \citet{lafreniere07, hinkley07, maroisADI, trident, masciadri05, biller04, schneider_roll, marois03, sparks02, marois00, racine99}. As AO systems continue to improve and eventually achieve Strehl ratios above $\sim$90\%, the light diffracted by the telescope aperture will become more important compared to scattered light and the use of a coronagraph will be mandatory. But even then, after removal of diffracted light by the coronagraph, high-contrast imaging applications will likely be limited by residual quasi-static speckles \citep[e.g.][]{macintosh_gpi}.

Many direct imaging searches for planetary or brown dwarf companions to stars have been done during the last five years, see for example \citet{biller07, chauvin06, metchev06, lowrance05, masciadri05, mccarthy04}; and \citet{luhman02} for searches carried out in $J$, $H$, or $K$, or \citet{kasper07} and \citet{heinze06} for searches made in $L^\prime$ or $M^\prime$. Depending on the observing strategy employed, the properties of the target stars, and the characteristics of the instrument used, each of these surveys was sensitive to a different regime of companion masses and separations. Typically, these surveys have reached detection contrasts of 10--13~mag for angular separations beyond 1\arcsec--2\arcsec, sufficient to detect planets more massive than $\sim$5~\mjup\ for targets aged $\sim$100~Myr. Unfortunately, rigorous statistical analyses allowing derivation of clear constraints on the population of planets in the regimes of mass and separation to which these surveys were sensitive are only beginning to be reported in the literature\footnote{In addition to the present work, analyses by \citet{nielsen07} and \citet{kasper07} have become available during the review process of this manuscript.}; an assessment of the current status of knowledge is thus rather difficult to make. Nonetheless, it is fair to say that the population of planets less massive than $\sim$5~\mjup, having orbits with a semi-major axis of tens to hundreds of AU, is poorly constrained.

In this paper we report the results of the Gemini Deep Planet Survey (GDPS), a direct imaging survey of 85 nearby young stars aimed at constraining the population of Jupiter mass planets with orbits of semi-major axis in the range 10-300~AU. The selection of the GDPS target sample is explained in \S\ref{sect:sample}, and the observations and data reduction are detailed in \S\ref{sect:obs}. The detection limits achieved for each target are then presented in \S\ref{sect:results} along with all candidate companions detected. A statistical analysis of the results allowing determination of the maximum fraction of stars that could bear planetary companions is presented in \S\ref{sect:discussion}. Concluding remarks follow in \S\ref{sect:conclusion}.

\section{Target sample}\label{sect:sample}

In light of the luminosity ratio and angular separation problem highlighted above, the list of target stars was assembled mainly on the basis of young age and proximity to the Sun, the latter yielding a larger angular separation for a given physical distance between the star and an eventual planet. Equivalently, a given detection threshold is achieved at a smaller physical separation for a star closer to the Sun, and planets on smaller orbits can be detected. Additionally, for angular separations where planet detection is limited by sky background noise or read noise, lower mass planets can be detected around a star closer to the Sun as their apparent brightness would be larger. Giant planets are intrinsically more luminous at young ages and fade with time (e.g. \citealp{marley07,baraffe03,burrows97}); therefore, for a given detection threshold, observations of younger stars are sensitive to planets having a lower mass. The proximity and age criteria used in building the target list thus maximize the range of mass and separation over which the survey is sensitive.

The target stars were selected from three sources: (1) Tables 3 and 4 of \citet{wichmann03}, which list nearby stars with an estimated age below or comparable to that of the Pleiades ($\sim$100~Myr), based on measurements of lithium abundance, space velocity, and X-ray activity; (2) Tables 2 and 5 of \citet{zuckerman04}, which list members of the $\beta$~Pictoris ($\sim$12~Myr) and AB~Doradus ($\sim$50~Myr) moving groups respectively; and (3) Tables 2 and 5 of \citet{montes01}, which list late-type single stars that are possible members of the Local Association (Pleiades moving group, 20--150~Myr) and IC~2391 supercluster (35--55~Myr) respectively, based on space velocity measurements. The stars listed in \citet{montes01} were initially selected based on various criteria indicative of youth, such as kinematic properties, rotation rate, chromospheric activity, lithium abundance, or X-ray emission, but for many of these stars the space velocity is the only indication of youth as other measurements are either unavailable or inconclusive; the young age of such stars is therefore uncertain. This uncertainty will be taken into account in our statistical analysis (\S\ref{sect:discussion}). A few stars known to have a circumstellar disk were added to these lists.

\clearpage
\onecolumngrid
\LongTables
\begin{landscape}
\begin{deluxetable}{llllrrcccccccccc}
\tablewidth{640pt}
\tabletypesize{\fontsize{5}{6}\selectfont}
\tablecolumns{16}
\tablecaption{GDPS target sample \label{tbl:targets}}
\tablehead{
\multicolumn{4}{c}{Names} & \colhead{$\alpha$} & \colhead{$\delta$} & \colhead{Spectral} & \colhead{H} & \colhead{Dist.\tablenotemark{a}} & \colhead{$\mu_\alpha\cos{\delta}$\tablenotemark{a}} & \colhead{$\mu_\delta$\tablenotemark{a}} & \colhead{[Fe/H]} & \colhead{[Fe/H]} & \colhead{Age} & \colhead{Age} & \colhead{Notes} \\
\cline{1-4}
\colhead{HD} & \colhead{GJ} & \colhead{HIP} & \colhead{Other} & \colhead{(J2000)} & \colhead{(J2000)} & \colhead{Type} & \colhead{(mag)} & \colhead{(pc)} & \colhead{(mas/yr)} & \colhead{(mas/yr)} & \colhead{} & \colhead{Ref.} & \colhead{(Myr)} & \colhead{Ref.} & \colhead{}
}
\startdata
166 & 5 & 544 & - & 00$^{\rm h}$06$^{\rm m}$36\fs78 & $+$29\degr01\arcmin17\farcs4 & K0 V & 4.63 & 13.7 & $  379.9$ & $ -178.3$ &  0.18 & V05 & 150--300 & G98,G00,F04,L06 & Her-Lyr \\
691 & - & 919 & V344 And & 00$^{\rm h}$11$^{\rm m}$22\fs44 & $+$30\degr26\arcmin58\farcs5 & K0 V & 6.26 & 34.1 & $  210.7$ & $   35.3$ &  0.32 & V05 & 50--280 & W03,W04,C05 & - \\
1405 & - & - & PW And & 00$^{\rm h}$18$^{\rm m}$20\fs90 & $+$30\degr57\arcmin22\farcs0 & K2 V & 6.51 & 30.6\tablenotemark{b} & $  143.7$\tablenotemark{b} & $ -171.5$\tablenotemark{b} & -0.78 & N04 & 50--50 & Z04b & AB Dor \\
5996 & - & 4907 & - & 01$^{\rm h}$02$^{\rm m}$57\fs22 & $+$69\degr13\arcmin37\farcs4 & G5 V & 5.98 & 25.8 & $  223.9$ & $ -148.4$ & -0.28 & K02 & 100--650 & M01a,G03 & ?LA \\
9540 & 59A & 7235 & - & 01$^{\rm h}$33$^{\rm m}$15\fs81 & $-$24\degr10\arcmin40\farcs7 & K0 V & 5.27 & 19.5 & $  271.9$ & $ -159.5$ & -0.02 & V05 & 100--1350 & M01a,W04 & ?LA \\
10008 & - & 7576 & - & 01$^{\rm h}$37$^{\rm m}$35\fs47 & $-$06\degr45\arcmin37\farcs5 & G5 V & 5.90 & 23.6 & $  171.0$ & $  -97.7$ & - & - & 150--300 & F04,L06 & Her-Lyr \\
- & 82 & 9291 & V596 Cas & 01$^{\rm h}$59$^{\rm m}$23\fs51 & $+$58\degr31\arcmin16\farcs1 & dM4e & 7.22 & 12.0 & $  321.8$ & $ -195.3$ & - & - & 50--50 & M06 & $\alpha$~Per \\
14802 & 97 & 11072 & kap For & 02$^{\rm h}$22$^{\rm m}$32\fs55 & $-$23\degr48\arcmin58\farcs8 & G2 V & 3.71 & 21.9 & $  197.3$ & $   -4.4$ & -0.04 & T05b & 5000--6700 & L99,W04,B99 & m \\
16765 & - & 12530 & 84 Cet & 02$^{\rm h}$41$^{\rm m}$14\fs00 & $-$00\degr41\arcmin44\farcs4 & F7 V & 4.64 & 21.6 & $  218.9$ & $ -125.3$ & -0.27 & N04 & 100--400 & M01a,H98,F95 & LA,m \\
17190 & 112 & 12926 & - & 02$^{\rm h}$46$^{\rm m}$15\fs21 & $+$25\degr38\arcmin59\farcs6 & K1 V & 6.00 & 25.7 & $  238.7$ & $ -149.9$ & -0.11 & V05 & 50--3500 & M01a,W04 & ?IC 2391 \\
17382 & 113 & 13081 & - & 02$^{\rm h}$48$^{\rm m}$09\fs14 & $+$27\degr04\arcmin07\farcs1 & K1 V & 5.69 & 22.4 & $  264.2$ & $ -127.8$ &  0.12 & T05b & 50--600 & M01a,W04 & ?IC 2391 \\
17925 & 117 & 13402 & EP Eri & 02$^{\rm h}$52$^{\rm m}$32\fs13 & $-$12\degr46\arcmin11\farcs0 & K2 V & 4.23 & 10.4 & $  398.1$ & $ -189.6$ &  0.18 & V05 & 40--128 & W03,M01a,M01b,L99 & LA \\
18803 & 120.2 & 14150 & 51 Ari & 03$^{\rm h}$02$^{\rm m}$26\fs03 & $+$26\degr36\arcmin33\farcs3 & G8 V & 5.02 & 21.2 & $  234.2$ & $ -168.2$ &  0.11 & V05 & 800--3600 & W04,T05a & - \\
19994 & 128 & 14954 & 94 Cet & 03$^{\rm h}$12$^{\rm m}$46\fs44 & $-$01\degr11\arcmin46\farcs0 & F8 V & 3.77 & 22.4 & $  193.4$ & $  -69.2$ &  0.19 & V05 & 800--3500 & W04,T05a & m \\
20367 & - & 15323 & - & 03$^{\rm h}$17$^{\rm m}$40\fs05 & $+$31\degr07\arcmin37\farcs4 & G0 V & 5.12 & 27.1 & $ -103.1$ & $  -56.6$ &  0.17 & S04a & 50--150 & W03 & - \\
- & - & - & 2E 759 & 03$^{\rm h}$20$^{\rm m}$49\fs50 & $-$19\degr16\arcmin10\farcs0 & K7 V & 7.66 & 27.0 & $   90.8$ & $  -43.8$ & - & - & 50--150 & M01a,L06 & LA \\
22049 & 144 & 16537 & eps Eri & 03$^{\rm h}$32$^{\rm m}$55\fs84 & $-$09\degr27\arcmin29\farcs7 & K2 V & 1.88 &  3.2 & $ -976.4$ & $   18.0$ & -0.03 & V05 & 530--930 & S00 & - \\
- & - & 17695 & - & 03$^{\rm h}$47$^{\rm m}$23\fs35 & $-$01\degr58\arcmin19\farcs9 & M3e & 7.17 & 16.3 & $  186.7$ & $ -271.8$ & - & - & 80--120 & L06 & LA (B4) \\
25457 & 159 & 18859 & - & 04$^{\rm h}$02$^{\rm m}$36\fs74 & $-$00\degr16\arcmin08\farcs1 & F6 V & 4.34 & 19.2 & $  151.2$ & $ -252.0$ &  0.02 & T05b & 80--120 & L06 & LA (B4) \\
283750 & 171.2 & 21482 & V833 Tau & 04$^{\rm h}$36$^{\rm m}$48\fs24 & $+$27\degr07\arcmin55\farcs9 & K2 V & 5.40 & 17.9 & $  232.4$ & $ -147.1$ & - & - & 50--150 & W03 & - \\
30652 & 178 & 22449 & 1 Ori & 04$^{\rm h}$49$^{\rm m}$50\fs41 & $+$06\degr57\arcmin40\farcs6 & F6 V & 1.76 &  8.0 & $  463.4$ & $   11.6$ &  0.03 & V05 & 50--500 & M01a,H99 & IC 2391 \\
- & 182 & 23200 & V1005 Ori & 04$^{\rm h}$59$^{\rm m}$34\fs83 & $+$01\degr47\arcmin00\farcs7 & M1 V & 6.45 & 26.7 & $   37.1$ & $  -93.9$ & - & - & 10--50 & F98,B99 & - \\
- & 234A & 30920 & V577 Mon & 06$^{\rm h}$29$^{\rm m}$23\fs40 & $-$02\degr48\arcmin50\farcs3 & M4 & 5.75 &  4.1 & $  694.7$ & $ -618.6$ & - & - & 100--3000 & M01a,M03 & ?LA,m \\
- & 281 & 37288 & - & 07$^{\rm h}$39$^{\rm m}$23\fs04 & $+$02\degr11\arcmin01\farcs2 & K7 & 6.09 & 14.9 & $ -147.9$ & $ -246.6$ & - & - & 150--300 & L06 & Her-Lyr \\
- & 285 & 37766 & YZ CMi & 07$^{\rm h}$44$^{\rm m}$40\fs17 & $+$03\degr33\arcmin08\farcs8 & M4.5 V & 6.01 &  5.9 & $ -344.9$ & $ -450.8$ & - & - & 50--50 & Z,M01 & LA \\
72905 & 311 & 42438 & 3 Uma & 08$^{\rm h}$39$^{\rm m}$11\fs70 & $+$65\degr01\arcmin15\farcs3 & G1.5 V & 4.28 & 14.3 & $  -27.7$ & $   87.9$ & -0.09 & T05b & 300--300 & S93,M01b & U~Ma \\
75332 & - & 43410 & - & 08$^{\rm h}$50$^{\rm m}$32\fs22 & $+$33\degr17\arcmin06\farcs2 & F7 Vn & 5.04 & 28.7 & $  -62.2$ & $  -85.0$ &  0.14 & V05 & 50--150 & W03 & - \\
77407 & - & 44458 & - & 09$^{\rm h}$03$^{\rm m}$27\fs08 & $+$37\degr50\arcmin27\farcs5 & G0 & 5.53 & 30.1 & $  -86.3$ & $ -168.8$ &  0.10 & V05 & 10--50 & W03,M01a,M01b & LA,m \\
78141 & - & - & - & 09$^{\rm h}$07$^{\rm m}$18\fs08 & $+$22\degr52\arcmin21\farcs6 & K0 V & 5.92 & 21.4\tablenotemark{c} & $   -0.4$\tablenotemark{d} & $  -67.6$\tablenotemark{d} & - & - & 50--150 & W03 & - \\
82558 & 355 & 46816 & LQ Hya & 09$^{\rm h}$32$^{\rm m}$25\fs57 & $-$11\degr11\arcmin04\farcs7 & K0 V & 5.60 & 18.3 & $ -248.2$ & $   35.1$ &  0.33 & V05 & 50--100 & W03,M01b & - \\
82443 & 354.1 & 46843 & DX Leo & 09$^{\rm h}$32$^{\rm m}$43\fs76 & $+$26\degr59\arcmin18\farcs7 & K0 V & 5.24 & 17.7 & $ -147.5$ & $ -246.3$ & -0.10 & T05b & 50--150 & W03,M01b & LA \\
- & 393 & 51317 & - & 10$^{\rm h}$28$^{\rm m}$55\fs55 & $+$00\degr50\arcmin27\farcs6 & M2 & 5.61 &  7.2 & $ -602.3$ & $ -731.9$ & - & - & 80--120 & L06 & LA (B4) \\
90905 & - & 51386 & - & 10$^{\rm h}$29$^{\rm m}$42\fs23 & $+$01\degr29\arcmin28\farcs0 & F5 & 5.60 & 31.6 & $ -151.4$ & $ -125.3$ &  0.07 & V05 & 50--150 & W03 & - \\
91901 & - & 51931 & - & 10$^{\rm h}$36$^{\rm m}$30\fs79 & $-$13\degr50\arcmin35\farcs8 & K2 V & 6.64 & 31.6 & $ -164.6$ & $   23.8$ & -0.03 & K02 & 50--5000 & M01a & ?IC 2391 \\
92945 & 3615 & 52462 & - & 10$^{\rm h}$43$^{\rm m}$28\fs27 & $-$29\degr03\arcmin51\farcs4 & K1 V & 5.77 & 21.6 & $ -215.4$ & $  -48.5$ &  0.13 & V05 & 80--120 & L06,W03,S04b,W04 & LA (B4) \\
93528 & - & 52787 & - & 10$^{\rm h}$47$^{\rm m}$31\fs16 & $-$22\degr20\arcmin52\farcs9 & K0 V & 6.56 & 34.9 & $ -122.7$ & $  -29.4$ &  0.11 & K02 & 50--150 & W03,S02 & - \\
- & 402 & 53020 & EE Leo & 10$^{\rm h}$50$^{\rm m}$52\fs06 & $+$06\degr48\arcmin29\farcs3 & M4 & 6.71 &  5.6 & $ -804.4$ & $ -809.6$ & - & - & 150--300 & L06 & Her-Lyr \\
96064 & - & 54155 & - & 11$^{\rm h}$04$^{\rm m}$41\fs47 & $-$04\degr13\arcmin15\farcs9 & G4 & 5.90 & 24.6 & $ -178.0$ & $ -104.1$ & -0.01 & T05b & 50--150 & W03 & m \\
97334 & 417 & 54745 & - & 11$^{\rm h}$12$^{\rm m}$32\fs35 & $+$35\degr48\arcmin50\farcs7 & G0 V & 5.02 & 21.7 & $ -248.6$ & $ -151.3$ &  0.09 & V05 & 80--300 & K01,M01a & ?LA \\
102195 & - & 57370 & - & 11$^{\rm h}$45$^{\rm m}$42\fs29 & $+$02\degr49\arcmin17\farcs3 & K0 V & 6.27 & 29.0 & $ -190.3$ & $ -111.4$ &  0.05 & S06 & 100--5000 & M01a & ?LA \\
102392 & - & 57494 & - & 11$^{\rm h}$47$^{\rm m}$03\fs83 & $-$11\degr49\arcmin26\farcs6 & K2 & 6.36 & 24.6 & $ -206.3$ & $  -60.7$ & - & - & 100--5000 & M01a & ?LA,m \\
105631 & 3706 & 59280 & - & 12$^{\rm h}$09$^{\rm m}$37\fs26 & $+$40\degr15\arcmin07\farcs4 & K0 V & 5.70 & 24.3 & $ -314.0$ & $  -51.3$ &  0.20 & V05 & 1600--1600 & M01b,W04 & - \\
107146 & - & 60074 & - & 12$^{\rm h}$19$^{\rm m}$06\fs50 & $+$16\degr32\arcmin53\farcs9 & G2 V & 5.61 & 28.5 & $ -175.7$ & $ -148.3$ & -0.03 & V05 & 50--100 & W03,Z04a & - \\
108767B & - & - & del Crv B & 12$^{\rm h}$29$^{\rm m}$51\fs85 & $-$16\degr30\arcmin55\farcs6 & K0 V & 6.37 & 26.9 & $ -210.0$ & $ -139.3$ & - & - & 40--260 & R05,G01,M01a & LA \\
109085 & 471.2 & 61174 & eta Crv & 12$^{\rm h}$32$^{\rm m}$04\fs23 & $-$16\degr11\arcmin45\farcs6 & F2 V & 3.37 & 18.2 & $ -424.4$ & $  -58.4$ & -0.05 & N04 & 600--1300 & M03,W05 & - \\
- & - & - & BD+60 1417 & 12$^{\rm h}$43$^{\rm m}$33\fs28 & $+$60\degr00\arcmin52\farcs7 & K0 & 7.36 & 17.7\tablenotemark{c} & $ -125.2$\tablenotemark{d} & $  -66.4$\tablenotemark{d} & - & - & 50--150 & W03 & - \\
111395 & 486.1 & 62523 & - & 12$^{\rm h}$48$^{\rm m}$47\fs05 & $+$24\degr50\arcmin24\farcs8 & G5 V & 4.71 & 17.2 & $ -334.6$ & $ -106.1$ &  0.13 & V05 & 600--1200 & H99,W04,T05a & - \\
113449 & - & 63742 & - & 13$^{\rm h}$03$^{\rm m}$49\fs65 & $-$05\degr09\arcmin42\farcs5 & G5 V & 5.67 & 22.1 & $ -189.8$ & $ -219.6$ & -0.22 & T05b & 80--120 & L06 & LA (B4) \\
- & 507.1 & 65016 & - & 13$^{\rm h}$19$^{\rm m}$40\fs12 & $+$33\degr20\arcmin47\farcs5 & M1.5 & 6.64 & 17.4 & $ -298.8$ & $ -143.9$ & - & - & 100--5000 & M01a & ?LA \\
116956 & - & 65515 & - & 13$^{\rm h}$25$^{\rm m}$45\fs53 & $+$56\degr58\arcmin13\farcs8 & G9 V & 5.48 & 21.9 & $ -217.4$ & $   11.2$ &  0.06 & T05b & 100--500 & M01a,G00 & LA \\
118100 & 517 & 66252 & EQ Vir & 13$^{\rm h}$34$^{\rm m}$43\fs21 & $-$08\degr20\arcmin31\farcs3 & K5 Ve & 6.31 & 19.8 & $ -287.4$ & $  -91.0$ &  0.00 & C97 & 50--50 & L05,M01a & ?LA \\
- & 524.1 & 67092 & - & 13$^{\rm h}$45$^{\rm m}$05\fs34 & $-$04\degr37\arcmin13\farcs2 & K5 & 7.33 & 25.7 & $ -159.9$ & $  -95.3$ & - & - & 100--5000 & M01a & ?LA \\
124106 & 3827 & 69357 & - & 14$^{\rm h}$11$^{\rm m}$46\fs17 & $-$12\degr36\arcmin42\farcs4 & K1 V & 5.95 & 23.1 & $ -257.5$ & $ -179.5$ & -0.10 & V05 & 800--1500 & H99,M01a,W04 & - \\
125161B & 9474B & - & - & 14$^{\rm h}$16$^{\rm m}$12\fs16 & $+$51\degr22\arcmin34\farcs7 & K1 & 6.32 & 29.8 & $ -150.0$ & $   89.4$ & - & - & 50--5000 & M01a & ?IC 2391 \\
129333 & 559.1 & 71631 & EK Dra & 14$^{\rm h}$39$^{\rm m}$00\fs21 & $+$64\degr17\arcmin30\farcs0 & G0 V & 6.01 & 33.9 & $ -138.6$ & $  -11.9$ &  0.16 & V05 & 20--120 & M01b,W03,L06 & LA (B4),m \\
130004 & - & 72146 & - & 14$^{\rm h}$45$^{\rm m}$24\fs18 & $+$13\degr50\arcmin46\farcs7 & K0 V & 5.67 & 19.5 & $ -232.9$ & $ -225.7$ & -0.24 & K02 & 100--5000 & M01a & ?LA \\
130322 & - & 72339 & - & 14$^{\rm h}$47$^{\rm m}$32\fs73 & $-$00\degr16\arcmin53\farcs3 & K0 V & 6.32 & 29.8 & $ -129.6$ & $ -140.8$ &  0.01 & V05 & 770--2400 & W04,S05 & - \\
130948 & 564 & 72567 & - & 14$^{\rm h}$50$^{\rm m}$15\fs81 & $+$23\degr54\arcmin42\farcs6 & G1 V & 4.69 & 17.9 & $  144.7$ & $   32.4$ &  0.05 & V05 & 50--150 & W03 & - \\
135363 & - & 74045 & - & 15$^{\rm h}$07$^{\rm m}$56\fs26 & $+$76\degr12\arcmin02\farcs7 & G5 & 6.33 & 29.4 & $ -131.9$ & $  169.3$ & -0.10 & K02 & 35--100 & W03,M01a,C05 & IC 2391,m \\
139813 & - & 75829  & - & 15$^{\rm h}$29$^{\rm m}$23\fs59 & $+$80\degr27\arcmin01\farcs0 & G5 & 5.56 & 21.7 & $ -217.4$ & $  105.5$ &  0.14 & V05 & 50--150 & W03 & - \\
141272 & 3917 & 77408 & - & 15$^{\rm h}$48$^{\rm m}$09\fs46 & $+$01\degr34\arcmin18\farcs3 & G8 V & 5.61 & 21.3 & $ -176.2$ & $ -166.7$ & -0.08 & T05b & 150--340 & G98,G00,W04,L06 & LA \\
147379B & 617B & 79762 & EW Dra & 16$^{\rm h}$16$^{\rm m}$45\fs31 & $+$67\degr15\arcmin22\farcs5 & M3 & 6.30 & 10.7 & $ -485.6$ & $   90.7$ &  0.00 & V04 & 100--1000 & M01a, H99 & ?LA \\
- & 628 & 80824 & V2306 Oph & 16$^{\rm h}$30$^{\rm m}$18\fs06 & $-$12\degr39\arcmin45\farcs3 & M3.5 & 5.37 &  4.3 & $  -93.6$ & $-1184.9$ & -0.25 & C01 & 100--5000 & M01a & LA \\
- & - & 81084 & - & 16$^{\rm h}$33$^{\rm m}$41\fs61 & $-$09\degr33\arcmin12\farcs0 & M0.5 & 7.78 & 31.9 & $  -67.4$ & $ -179.7$ & - & - & 80--120 & L06 & LA (B4) \\
160934 & 4020A & 86346 & - & 17$^{\rm h}$38$^{\rm m}$39\fs63 & $+$61\degr14\arcmin16\farcs1 & K7 & 7.00 & 24.5 & $  -31.2$ & $   59.4$ & - & - & 30--50 & Z04c,L06 & AB Dor,m \\
162283 & 696 & 87322 & - & 17$^{\rm h}$50$^{\rm m}$34\fs03 & $-$06\degr03\arcmin01\farcs0 & M0 & 6.70 & 21.9 & $  -26.1$ & $ -131.4$ & - & - & 100--5000 & M01a, B98 & - \\
166181 & - & 88848 & V815 Her & 18$^{\rm h}$08$^{\rm m}$16\fs03 & $+$29\degr41\arcmin28\farcs1 & G6 V & 5.76 & 32.3\tablenotemark{e} & $  107.0$\tablenotemark{e} & $  -31.0$\tablenotemark{e} & -0.70 & N04 & 50--150 & W03 & m \\
167605 & - & 89005 & LP Dra & 18$^{\rm h}$09$^{\rm m}$55\fs50 & $+$69\degr40\arcmin49\farcs8 & K2 V & 6.46 & 31.0 & $  -25.3$ & $  193.9$ &  0.13 & K02 & 50--5000 & M01a & ?IC 2391,m \\
187748 & - & 97438 & - & 19$^{\rm h}$48$^{\rm m}$15\fs45 & $+$59\degr25\arcmin22\farcs4 & G0 & 5.32 & 28.4 & $   15.8$ & $  116.5$ & -0.06 & N04 & 50--150 & W03 & - \\
- & 791.3 & 101262 & - & 20$^{\rm h}$31$^{\rm m}$32\fs07 & $+$33\degr46\arcmin33\farcs1 & K5 V & 6.64 & 26.2 & $  142.1$ & $   16.6$ & - & - & 50--1000 & M01a,H99 & ?IC 2391 \\
197481 & 803 & 102409 & AU Mic & 20$^{\rm h}$45$^{\rm m}$09\fs53 & $-$31\degr20\arcmin27\farcs2 & M0 & 4.83 &  9.9 & $  280.4$ & $ -360.1$ & - & - & 8--20 & Z01 & $\beta$~Pic \\
201651 & - & 104225 & - & 21$^{\rm h}$06$^{\rm m}$56\fs39 & $+$69\degr40\arcmin28\farcs5 & K0 & 6.41 & 32.8 & $  109.0$ & $   66.0$ & -0.18 & K02 & 50--5900 & M01a,W04 & ?IC 2391 \\
202575 & 824 & 105038 & - & 21$^{\rm h}$16$^{\rm m}$32\fs47 & $+$09\degr23\arcmin37\farcs8 & K3 V & 5.53 & 16.2 & $  146.5$ & $ -119.1$ &  0.04 & V05 & 100--1000 & M01a,H99 & ?LA \\
- & 4199 & 106231 & LO Peg & 21$^{\rm h}$31$^{\rm m}$01\fs71 & $+$23\degr20\arcmin07\farcs4 & K8 & 6.52 & 25.1 & $  134.1$ & $ -144.8$ & - & - & 30--50 & Z04c,L06 & AB~Dor \\
206860 & 836.7 & 107350 & HN Peg & 21$^{\rm h}$44$^{\rm m}$31\fs33 & $+$14\degr46\arcmin19\farcs0 & G0 V & 4.60 & 18.4 & $  231.1$ & $ -113.5$ & -0.02 & V05 & 150--300 & L06,G98,G00 & Her-Lyr \\
208313 & 840 & 108156 & - & 21$^{\rm h}$54$^{\rm m}$45\fs04 & $+$32\degr19\arcmin42\farcs9 & K0 V & 5.68 & 20.3 & $  210.6$ & $ -233.4$ & -0.04 & V05 & 100--1000 & M01a,H99 & ?LA \\
- & - & - & V383 Lac & 22$^{\rm h}$20$^{\rm m}$07\fs03 & $+$49\degr30\arcmin11\farcs8 & K1 V & 6.58 & 27.5\tablenotemark{b} & $   93.4$\tablenotemark{b} & $    5.0$\tablenotemark{b} & - & - & 50--150 & W03,M01a,C05 & LA \\
213845 & 863.2 & 111449 & ups Aqr & 22$^{\rm h}$34$^{\rm m}$41\fs64 & $-$20\degr42\arcmin29\farcs6 & F7 V & 4.27 & 22.7 & $  221.6$ & $ -146.6$ &  0.11 & T05b & 150--300 & L06 & Her-Lyr,m \\
- & 875.1 & 112909 & GT Peg & 22$^{\rm h}$51$^{\rm m}$53\fs54 & $+$31\degr45\arcmin15\farcs2 & M3 & 7.13 & 14.2 & $  527.0$ & $  -50.6$ & - & - & 200--300 & L05,M01a & ?IC 2391 \\
- & 876 & 113020 & IL Aqr & 22$^{\rm h}$53$^{\rm m}$16\fs73 & $-$14\degr15\arcmin49\farcs3 & M4 & 5.35 &  4.7 & $  960.3$ & $ -675.6$ & - & - & 100--5000 & M01a & ?LA \\
- & 9809 & 114066 & - & 23$^{\rm h}$06$^{\rm m}$04\fs84 & $+$63\degr55\arcmin34\farcs4 & M0 & 7.17 & 24.9 & $  171.0$ & $  -58.5$ & - & - & 30--50 & Z04c,L06 & AB~Dor \\
220140 & - & 115147 & V368 Cep & 23$^{\rm h}$19$^{\rm m}$26\fs63 & $+$79\degr00\arcmin12\farcs7 & K1 V & 5.51 & 19.7 & $  201.3$ & $   71.6$ & -0.64 & N04 & 50--150 & M01b,W03 & LA,m \\
221503 & 898A & 116215 & - & 23$^{\rm h}$32$^{\rm m}$49\fs40 & $-$16\degr50\arcmin44\farcs3 & K5 & 5.61 & 13.9 & $  343.6$ & $ -217.7$ &  0.00 & C04 & 100--800 & M01a,H99 & ?LA \\
- & 900 & 116384 & - & 23$^{\rm h}$35$^{\rm m}$00\fs28 & $+$01\degr36\arcmin19\farcs5 & K7 & 6.28 & 19.3 & $  343.0$ & $   26.8$ & -0.10 & C01 & 150--250 & Z06 & Ca-Near,m \\
- & 907.1 & 117410 & - & 23$^{\rm h}$48$^{\rm m}$25\fs69 & $-$12\degr59\arcmin14\farcs8 & K8 & 6.49 & 27.1 & $  233.6$ & $   22.1$ & - & - & 150--250 & Z06 & Ca-Near,m \\
\enddata
\tablenotetext{a}{From the Hipparcos catalog \citep{hipparcos}, unless stated otherwise.}
\tablenotetext{b}{From \citet{montes01}.}
\tablenotetext{c}{From the Tycho catalog \citep{tycho}.}
\tablenotetext{d}{From the Tycho-2 catalog \citep{tycho2}.}
\tablenotetext{e}{From \citet{fekel05}.}
\tablecomments{Star is a member of ($\alpha$~Per) $\alpha$~Persei; (AB~Dor) AB~Doradus; ($\beta$~Pic) $\beta$~Pictoris; (Ca-Near) Carina-Near; (Her-Lyr) Hercules-Lyra; (LA) Local association; (LA (B4)) Local association, subgroup B4; (U~Ma) Ursa Major. If a question mark precedes the association, the membership is doubtful or based on kinematics only. An ``m'' indicates that the star is a multiple, see \S~\ref{sect:binaries} for more detail.}
\tablerefs{(Z) B. Zuckerman, private communication.; (B99) \citealp{barrado99}; (C05) \citealp{carpenter05}; (C01) \citealp{cayrel01}; (C97) \citealp{cayrel97}; (C04) \citealp{clem04}; (F95) \citealp{favata95}; (F98) \citealp{favata98}; (F04) \citealp{fuhrmann04}; (G00) \citealp{gaidos00}; (G98) \citealp{gaidos98}; (G01) \citealp{gerbaldi01}; (G03) \citealp{gray03}; (H98) \citealp{huensch98}; (H99) \citealp{huensch99}; (K01) \citealp{kirkpatrick01}; (K02) \citealp{kotoneva02}; (L99) \citealp{lachaume99}; (L06) \citealp{lopez-santiago06}; (L05) \citealp{lowrance05}; (M06) \citealp{makarov06}; (M03) \citealp{mohanty03}; (M01a) \citealp{montes01}; (M01b) \citealp{montes01b}; (N04) \citealp{nordstrom04}; (R05) \citealp{rieke05}; (S05) \citealp{saffe05}; (S04a) \citealp{santos04}; (S93) \citealp{soderblom93}; (S00) \citealp{song00}; (S04b) \citealp{song04}; (S06) \citealp{sousa06}; (T05a) \citealp{takeda05}; (T05b) \citealp{taylor05}; (V04) \citealp{valdes04}; (V05) \citealp{valenti05}; (W03) \citealp{wichmann03}; (W04) \citealp{wright04}; (W05) \citealp{wyatt05}; (Z01) \citealp{zuckerman01}; (Z04b) \citealp{zuckerman04}; (Z04a) \citealp{zuckerman04b}; (Z04c) \citealp{zuckerman04c}; (Z06) \citealp{zuckerman06}}
\end{deluxetable}
\clearpage
\end{landscape}
\twocolumngrid

From this preliminary compilation, we have retained only stars with a distance smaller than 35~pc, and we have excluded stars of declination below $-32\degr$ since observations were to be made from the Gemini North observatory. Finally, we have further excluded stars indicated to be multiple in \citet{zuckerman04}. This procedure yielded a list of slightly over 100 target stars, of which 85 were actually observed. The properties of these 85 stars are presented in Table~\ref{tbl:targets} and Figure~\ref{fig:targets}. The median spectral type of our sample is K0, the median $H$ magnitude is 5.75, the median distance is 22~pc, the median proper motion amplitude is 240~mas~yr$^{-1}$, and the median [Fe/H] is 0.00~dex (standard deviation of 0.21~dex).

Despite our effort to select only single stars, our observations show that 16 of the 85 target stars are close double or triple systems; this is indicated in the last column of Table~\ref{tbl:targets}. A thorough review of the literature revealed that 11 of these were known at the time the target list was compiled, two of which are astrometric multiples that had never been resolved prior to our observations (HD~14802 and HD~166181). Five other multiple systems were resolved with AO only after the target list was compiled (HD~77407, HD~129333, HD~135363, HD~160934, and HD~220140). Finally, the star HD~213845 is reported to be part of a binary system for the first time here. The multiple systems observed are discussed further in \S\ref{sect:binaries}.

Age estimates for the stars in our sample, needed to convert the observed contrasts into mass detection limits using evolution models of giant planets,\footnote{It is assumed that any planet and its primary star would be coeval.} are reported in Table~\ref{tbl:targets} along with the references used for their determination. Whenever possible, we have used ages stated explicitly in the literature or the age of the association to which a star belongs. When no specific age estimate was available for stars taken from \citet{wichmann03}, ages of 10--50~Myr or 50--150~Myr were assigned to the stars having a lithium abundance above or comparable to that of the Pleiades, respectively. For other stars that have lithium and/or X-ray measurements, ages were estimated from a comparison of the Li~I~6708~\AA\ equivalent width and/or the ratio of the X-ray to bolometric luminosity with Figures~3 and/or 4 of \citet{zuckerman04} respectively. When lithium or X-ray measurements were not available, the kinematic ages were used as lower limits while the ages derived from the chromospheric activity index, $\log{R_{\rm HK}}$, were used as upper limits, as \citet{song04} showed that the latter ages tend to be systematically higher than those derived from lithium abundance or X-ray emission. When only the value of $\log{R_{\rm HK}}$ was available, the calibration of \citet{donahue93}\footnote{This calibration is given explicitly in \citet{henry96}.} was used to obtain an age estimate. Finally, when only kinematics measurements were available for a given star, an age of 100--5000~Myr or 50--5000~Myr was assigned if the star is a possible member of the Local Association or the IC~2391 supercluster respectively.

\section{Observations and image processing}\label{sect:obs}

\subsection{Data acquisition and observing strategy}

All observations were obtained at the Gemini North telescope with the Altair adaptive optics system \citep{herriot00} and the NIRI camera \citep{hodapp03} (programs GN-2004B-Q-14, GN-2005A-Q-16, GN-2005B-Q-4, GN-2006A-Q-5, and GN-2006B-Q-5). The $f/32$ camera was used, yielding 0.022\arcsec\ pixel$^{-1}$ and a field of view of $22\arcsec\times22\arcsec$. The field lens of Altair, which improves the off-axis adaptive optics correction, was not used for any observation as it was not available for the first epoch observations. Because it introduces an undetermined field distortion, having used the field lens for the second epoch observations only would have complicated or prevented verification of the physical association of companion candidates identified in the first epoch observations. The observations were obtained in the narrow band filter CH4-short (1.54--1.65~$\micron$), for the following reason. According to evolution models (e.g. \citealp{baraffe03}), planetary mass objects older than 10-20~Myr should have an effective temperature below 1000~K. Because of the large amounts of methane and the increased collision induced absorption by H$_2$ in their atmosphere, the near-infrared $K$-band flux of such objects is largely suppressed. It is thus more efficient to search for giant planets in either the $J$ or the $H$ band; the latter was preferred in this study because higher Strehl ratios are achieved at longer wavelengths. As the bulk of the $H$-band flux of cool giant planets is emitted in a narrow band centered at $\sim$1.58~\micron\ because of important absorption by methane beyond 1.6~$\mu$m, it is even more efficient to search for these planets using the CH4-short filter, which is well matched to the peak of the emission. Based on evolution models and synthetic spectra of giant planets \citep{baraffe03}, it is expected that the mean flux density of a planet in the NIRI CH4-short filter be between 1.5 and 2.5 times higher (0.44--1.0~mag brighter) than in the broad band $H$ filter, depending on the specific age and mass of the planet. These factors are consistent with the factors 1.6-2.0 (0.5--0.75~mag) calculated from the observed spectra of T7--T8 brown dwarfs, which have $T_{\rm eff}\sim800$~K.

The angular differential imaging (ADI, \citealp{maroisADI}) technique was used to suppress the PSF speckle noise and improve our sensitivity to faint companions. This technique consists of acquiring a sequence of many exposures of the target using an altitude/azimuth telescope with the instrument rotator turned off (at the Cassegrain focus) to keep the instrument and telescope optics aligned. This is a very stable configuration and ensures a high correlation of the sequence of PSF images. This setup also causes a rotation of the field of view (FOV) during the sequence. For each target image in such a sequence, it is possible to build a reference image from other target images in which any companion would be sufficiently displaced due to FOV rotation. After subtraction of the reference image, the residual images are rotated to align their FOV and co-added. Because of the rotation, the residual PSF speckle noise is averaged incoherently, ensuring an ever improving detection limit with increasing exposure time. It has been shown that, for ADI with Altair/NIRI, the subtraction of an optimized reference PSF image from a target image can suppress the PSF speckle noise by a factor of $\sim$12, and that a noise suppression factor of $\sim$100 is achieved for the combination of 90 such difference images \citep{lafreniere07, maroisADI}.

An individual exposure time of 30 seconds was chosen for all targets. This exposure time is long enough so that, at large separation, faint companion detection is limited by sky background noise rather than read noise, and short enough so that the radius of the region affected by saturation and non-linearity of the detector typically does not exceed 0.5\arcsec. The nominal observing sequence consisted of 90 images, but oftentimes a few images had to be discarded due to brief periods of very bad seeing, loss of tracking, or the advent of clouds. No dithering was made during the main observing sequence to ensure a high correlation of the PSF images; flat-\\
\onecolumngrid
\begin{deluxetable}{lccccc}
\tablewidth{0pt}
\tabletypesize{\scriptsize}
\tablecolumns{6}
\tablecaption{GDPS observation log \label{tbl:obs}}
\tablehead{
\colhead{Name} & \colhead{Date} & \colhead{Number of} & \colhead{Strehl} & \colhead{FOV rotation} & \colhead{Saturation} \\
\colhead{} & \colhead{} & \colhead{exposures} & \colhead{(\%)} & \colhead{(deg)} & \colhead{radius (\arcsec)\tablenotemark{a}}
}
\startdata
HD 166 & 2005/08/25 &  83 & 5-8 &  55 & 0.98 \\
 & 2006/07/18 &  83 & 7-10 &  81 & 0.78 \\
HD 691 & 2005/08/10 &  90 & 13-17 &  70 & 0.43 \\
 & 2006/09/18 & 117 & 16-30 &  88 & 0.44 \\
HD 1405 & 2004/08/22 &  90 & 4-10 &  17 & 0.53 \\
 & 2005/08/04 &  90 & 6-18 &  69 & 0.40 \\
HD 5996 & 2005/08/12 &  90 & 18-20 &  24 & 0.50 \\
 & 2006/09/25 &  90 & 15-17 &  21 & 0.50 \\
HD 9540 & 2005/08/14 &  90 & 16-19 &  25 & 0.55 \\
 & 2006/09/28 &  45 & 14-17 &  11 & 0.61 \\
HD 10008 & 2005/08/10 &  90 & 18-20 &  36 & 0.51 \\
GJ 82 & 2005/08/31 &  90 & 10-12 &  27 & 0.28 \\
HD 14802 & 2005/08/20 &  90 & - &  23 & 1.09 \\
HD 16765 & 2005/09/10 &  90 & 14-17 &  45 & 0.72 \\
HD 17190 & 2005/08/24 &  90 & 13-30 & 108 & 0.52 \\
HD 17382 & 2004/12/22 &  66 & 15 &  68 & 0.55 \\
 & 2005/09/11 &  90 & 19-23 & 104 & 0.52 \\
HD 17925 & 2004/11/04 &  83 & ${\scriptstyle \gtrsim}$18\tablenotemark{b} &  29 & 0.66 \\
HD 18803 & 2004/12/24 &  90 & 7-14 &  99 & 0.70 \\
 & 2005/09/12 &  78 & 17-18 & 108 & 0.64 \\
HD 19994 & 2005/08/31 &  90 & - &  44 & 0.83 \\
 & 2006/10/01 &  57 & - &  27 & 0.72 \\
HD 20367 & 2005/10/02 &  90 & 12-14 &  67 & 0.70 \\
2E 759 & 2005/10/17 &  59 & 7-10 &  31 & 0.22 \\
HD 22049 & 2005/09/08 &  90 & - &  32 & 2.06 \\
HIP 17695 & 2005/09/13 &  89 & 20-20 &  45 & 0.24 \\
HD 25457 & 2005/10/02 &  90 & - &  43 & 0.96 \\
HD 283750 & 2004/10/24 &  90 & 15 &  99 & 0.54 \\
 & 2005/10/04 &  87 & 19-23 & 101 & 0.59 \\
HD 30652 & 2005/09/12 &  52 & - &  35 & 1.86 \\
GJ 182 & 2004/11/05 &  90 & 16-20 &  31 & 0.37 \\
 & 2005/10/17 &  33 & 11-11 &  29 & 0.39 \\
GJ 234A & 2005/11/05 &  72 & 16 &  34 & 0.42 \\
GJ 281 & 2005/03/25 &  67 & 9-10 &  49 & 0.52 \\
 & 2006/02/12 &  25 & 8-9 &  11 & 0.47 \\
GJ 285 & 2005/03/18 &  20 & - &  10 & 0.45 \\
 & 2006/02/12 &  90 & 4-5 &  73 & 0.55 \\
HD 72905 & 2005/04/23 &  84 & 7 &  25 & 0.87 \\
HD 75332 & 2005/04/24 &  89 & ${\scriptstyle \gtrsim}$17\tablenotemark{b} &  27 & 0.59 \\
 & 2006/12/20 &  16 & ${\scriptstyle \gtrsim}$16\tablenotemark{b} &  11 & 0.59 \\
HD 77407 & 2005/04/26 &  84 & 16-19 &  33 & 0.61 \\
HD 78141 & 2004/12/21 &  85 & 14-16 &  19 & 0.55 \\
HD 82558 & 2005/04/18 &  90 & - &  30 & 0.61 \\
HD 82443 & 2004/12/25 &  75 & 18 &  28 & 0.61 \\
GJ 393 & 2005/04/20 &  90 & 13-15 &  44 & 0.55 \\
HD 90905 & 2005/03/18 &  90 & 13-18 &  47 & 0.61 \\
 & 2006/04/11 &  35 & 13-15 &  14 & 0.55 \\
HD 91901 & 2005/04/29 &  71 & 9 &  22 & 0.44 \\
HD 92945 & 2005/05/26 &  85 & 15-16 &  19 & 0.61 \\
 & 2006/05/16 &  10 & 10-11 &   2 & 0.53 \\
HD 93528 & 2005/04/30 &  86 & - &  26 & 0.39 \\
GJ 402 & 2005/04/26 &  79 & 12-16 &  37 & 0.39 \\
 & 2006/02/16 &  60 & 6-10 &  33 & 0.35 \\
HD 96064 & 2005/04/19 &  89 & 21-23 &  37 & 0.50 \\
 & 2006/03/05 &  90 & 13-19 &  36 & 0.50 \\
HD 97334 & 2005/04/18 &  90 & 16-17 &  54 & 0.70 \\
HD 102195 & 2005/04/24 &  91 & 20-21 &  54 & 0.41 \\
 & 2006/03/18 &  82 & 12-18 &  30 & 0.39 \\
HD 102392 & 2005/04/23 &  89 & 19-24 &  32 & 0.39 \\
 & 2006/03/12 &  90 & 9-13 &  31 & 0.40 \\
HD 105631 & 2005/05/29 &  90 & 14-19 &  45 & 0.55 \\
HD 107146 & 2005/05/30 &  90 & 21-26 &  71 & 0.57 \\
HD 108767B & 2005/04/22 &  90 & 14 &  27 & 0.43 \\
 & 2006/02/16 &  43 & 10-11 &  14 & 0.41 \\
HD 109085 & 2005/05/26 &  90 & - &  22 & 1.09 \\
 & 2006/03/12 &  15 & - &   3 & 1.09 \\
BD+60 1417 & 2005/04/18 &  90 & 18-23 &  24 & 0.26 \\
 & 2006/04/11 &  63 & 12 &  19 & 0.24 \\
HD 111395 & 2005/04/19 &  89 & ${\scriptstyle \gtrsim}$12\tablenotemark{b} & 120 & 0.77 \\
HD 113449 & 2005/06/01 &  47 & 10-20 &  37 & 0.52 \\
GJ 507.1 & 2005/06/07 &  87 & 5-7 &  61 & 0.44 \\
HD 116956 & 2005/05/29 &  90 & 5-14 &  27 & 0.55 \\
 & 2006/05/16 &  60 & 5-8 &  18 & 0.61 \\
HD 118100 & 2005/04/27 &  53 & - &  18 & 0.39 \\
GJ 524.1 & 2005/04/18 &  90 & 18-25 &  37 & 0.26 \\
 & 2006/05/18 &  90 & 13-14 &  37 & 0.22 \\
HD 124106 & 2005/04/19 &  86 & 18-19 &  32 & 0.50 \\
 & 2006/02/16 &  80 & 10-12 &  24 & 0.50 \\
HD 125161B & 2005/05/30 &  90 & 17-23 &  31 & 0.39 \\
HD 129333 & 2005/04/20 &  90 & 19-20 &  22 & 0.48 \\
HD 130004 & 2005/05/25 &  90 & 17-18 & 105 & 0.55 \\
HD 130322 & 2005/05/27 &  88 & 15-19 &  40 & 0.48 \\
 & 2006/05/15 &  10 & 10-10 &   5 & 0.40 \\
HD 130948 & 2005/04/17 &  90 & ${\scriptstyle \gtrsim}$9\tablenotemark{b} & 122 & 0.83 \\
HD 135363 & 2005/04/18 &  87 & 14-15 &  19 & 0.48 \\
 & 2006/02/16 &  60 & 8-9 &  14 & 0.44 \\
HD 139813 & 2005/05/30 &  90 & ${\scriptstyle \gtrsim}$12\tablenotemark{b} &  20 & 0.57 \\
HD 141272 & 2005/04/19 &  90 & 18-19 &  47 & 0.55 \\
 & 2006/03/12 &  42 & 13 &  20 & 0.56 \\
HD 147379B & 2005/04/18 &  90 & 17-17 &  22 & 0.50 \\
GJ 628 & 2005/04/17 &  90 & 11 &  29 & 0.70 \\
 & 2006/04/11 &  40 & 9-14 &  13 & 0.66 \\
HIP 81084 & 2005/04/19 &  73 & 17-18 &  30 & 0.33 \\
 & 2006/05/15 &  90 & 8-13 &  31 & 0.22 \\
HD 160934 & 2005/04/18 &  84 & 17-24 &  24 & 0.35 \\
 & 2006/09/17 &  14 & 12-14 &   4 & 0.34 \\
HD 162283 & 2005/04/20 & 120 & 15-19 &  45 & 0.38 \\
 & 2006/09/16 & 100 & 27-29 &  31 & 0.33 \\
HD 166181 & 2005/04/17 &  90 & 16 &  76 & 0.59 \\
 & 2006/09/18 &  45 & 18-21 &  37 & 0.48 \\
HD 167605 & 2005/05/27 &  90 & 20 &  22 & 0.39 \\
HD 187748 & 2005/05/25 &  97 & 15-19 &  30 & 0.66 \\
 & 2006/09/15 &  75 & ${\scriptstyle \gtrsim}$22\tablenotemark{b} &  21 & 0.50 \\
GJ 791.3 & 2005/05/26 &  87 & 9-19 &  54 & 0.42 \\
HD 197481 & 2005/07/29 &  68 & 6-10 &  21 & 0.87 \\
HD 201651 & 2005/06/27 &  90 & 18-23 &  21 & 0.38 \\
 & 2006/09/14 &  30 & 19-21 &   7 & 0.38 \\
HD 202575 & 2005/07/16 &  90 & 17-23 &  75 & 0.57 \\
 & 2006/09/14 &  30 & 16-18 &   9 & 0.56 \\
GJ 4199 & 2004/08/23 &  65 & 10-13 & 118 & 0.44 \\
 & 2005/08/04 &  90 & 15-23 & 136 & 0.39 \\
HD 206860 & 2005/08/10 &  34 & ${\scriptstyle \gtrsim}$13\tablenotemark{b} &  56 & 0.77 \\
 & 2006/06/26 &  60 & ${\scriptstyle \gtrsim}$15\tablenotemark{b} &  80 & 0.61 \\
HD 208313 & 2005/06/27 &  90 & 23-23 &  67 & 0.46 \\
 & 2006/06/25 &  89 & 14-22 &  66 & 0.55 \\
V383 Lac & 2005/07/26 &  66 & 13-17 &  28 & 0.42 \\
 & 2006/06/30 &  77 & 15-18 &  27 & 0.32 \\
HD 213845 & 2005/08/24 &  90 & - &  26 & 0.81 \\
 & 2006/07/06 &  90 & - &  24 & 0.83 \\
GJ 875.1 & 2005/08/10 &  90 & 16-18 &  69 & 0.33 \\
 & 2006/07/07 &  79 & 7-17 &  61 & 0.31 \\
GJ 876 & 2005/08/21 &  82 & 9-16 &  28 & 0.68 \\
GJ 9809 & 2005/08/04 &  90 & 18-20 &  25 & 0.31 \\
 & 2006/09/14 & 120 & 25-27 &  31 & 0.22 \\
HD 220140 & 2005/08/05 &  90 & 16-18 &  21 & 0.59 \\
 & 2006/07/16 &  82 & 7-9 &  19 & 0.63 \\
HD 221503 & 2005/08/31 &  90 & 21-22 &  28 & 0.52 \\
GJ 900 & 2004/08/24 &  90 & 15-21 &  17 & 0.46 \\
 & 2005/09/08 &  90 & 16-22 &  46 & 0.42 \\
GJ 907.1 & 2005/09/07 &  65 & 5-15 &  22 & 0.37 \\
 & 2006/07/17 &  44 & 8 &  16 & 0.31
\enddata
\tablenotetext{a}{Radius at which the PSF radial intensity profile reaches 75\% of the detector well capacity.}
\tablenotetext{b}{Only a lower estimate of the Strehl ratio can be obtained as the PSF peak is in the non-linear regime or sligthly saturated.}
\end{deluxetable}
\twocolumngrid

\begin{figure*}
\epsscale{0.95}
\plotone{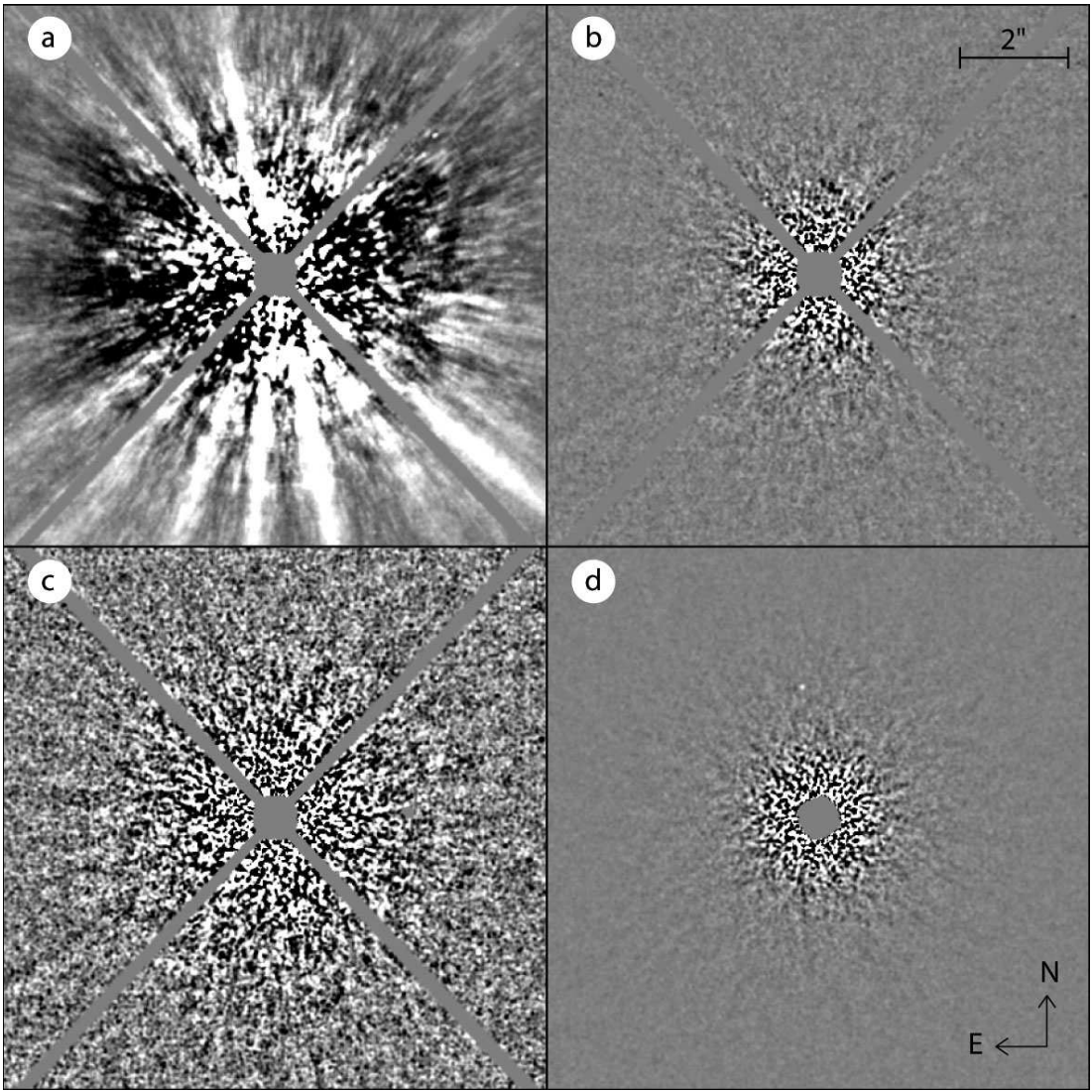}
\caption{\label{fig:im_sub} Illustration of the ADI noise attenuation process. Panel (a) shows an original 30-s image of the young star HD~691 after subtraction of an azimuthally symmetric median intensity profile, panels (b) and (c) both show, with a different intensity scale, the corresponding residual image after ADI subtraction using the LOCI algorithm, and panel (d) shows the median combination of 117 such residual images. Display intensity ranges are $\pm5\times10^{-6}$ and $\pm10^{-6}$ of stellar PSF peak for the top and bottom rows respectively. Each panel is 10\arcsec\ on a side. The diffraction spikes from the secondary mirror support vanes and the central saturated region are masked. The faint point source ($\Delta m=14.9$) visible in panel (d) at a separation of 2.43\arcsec\ and P.A. of 7.3\degr\ could not have been detected without ADI processing.}
\end{figure*}
\noindent field errors, bad pixels, and cosmic ray hits are naturally averaged/removed with ADI because of the FOV rotation. The PSF centroid was found to wander over the detector by typically 2-5 pixels throughout an observing sequence because of mechanical flexure and differential refraction between the wavefront sensing and science wavelengths; for a handful of targets the variation slightly exceeded 10 pixels. Short unsaturated exposures were acquired before and after the main sequence of (saturated) images for photometric calibration and Strehl ratio estimation; these observations were acquired in sub-array mode ($256\times256$ or $512\times512$ pixels), for which the minimum exposure time is shorter. Typically, an unsaturated sequence consisted of five exposures each obtained at a different dither position. The unsaturated observations are missing for a few targets as they were either skipped in the execution of the program, or they turned out to be saturated despite using the shortest possible exposure time.  Table~\ref{tbl:obs} summarizes all observations. The last column of the table (``saturation radius'') indicates the separation at which the radial profile of the PSF reaches 75\% of the detector full well capacity; linearity should be better than 1\% at this level \citep{hodapp03}. We have not analyzed the data inside this separation; point sources located at least one PSF full-width-at-half-maximum (FWHM) past this separation can be detected in our analysis, provided that their brigthness is above the detection limit.

\subsection{Data reduction}\label{sect:datared}

For each sequence of short unsaturated exposures, a sky frame was constructed by taking the median of the images obtained at different dither positions; this sky frame was subtracted from each image. The images were then divided by a flat field image. The PSFs of a given unsaturated sequence were registered to a common center and the median of the image sequence was obtained. The center of the PSFs were determined by fitting a 2-dimensional Gaussian function. As an indication of the quality of an observing sequence, the Strehl ratio was calculated by comparing the peak pixel value of the observed PSF image with that of an appropriate theoretical PSF. The calculated Strehl ratio values are reported in Table~\ref{tbl:obs}; two values are indicated for a target when unsaturated data were obtained before {\it and} after the main saturated sequence. Strehl ratios were typically in the range 10--20\%.

Images of the main saturated sequence were first divided by a flat field image. Bad and hot pixels, as determined from analysis of the flat field image and dark frame respectively, were replaced by the median value of neighboring pixels. Field distortion was corrected using an IDL procedure provided by the Gemini staff (C. Trujillo, private communication) and modified to use the IDL {\sl interpolate} function with cubic interpolation. The plate scale and field of view orientation for each image were obtained from the FITS header keywords.

For each sequence of saturated images, the stellar PSF of the first image was registered to the image center by maximizing the cross-correlation of the PSF diffraction spikes with themselves in a 180-degree rotation of the image about its center. The stellar PSF of the subsequent images was registered to the image center by maximizing the cross-correlation of the PSF diffraction spikes with those in the first image. Prior to shifting, the $1024\times1024$~pixel images were padded with zeros to $1450\times1450$~pixel to ensure that no FOV would be lost. An azimuthally symmetric intensity profile was finally subtracted from each image to remove the smooth seeing halo.

Next, the stellar PSF speckles were removed from each image by subtracting an optimized reference PSF image obtained using the ``locally optimized combination of images'' (LOCI) algorithm detailed in \citet{lafreniere07}. The heart of this algorithm consists in dividing the target image into subsections and obtaining, for each subsection independently, an optimized reference PSF image consisting of a linear combination of the other images of the sequence for which the rotation of the FOV would have displaced sufficiently an eventual companion. For each subsection, the coefficients of the linear combination are optimized such that its subtraction from the target image minimizes the noise.  The subsections geometry and the algorithm parameters determined in \citet{lafreniere07} were used for all targets. The residual images were then rotated to align their FOV and their median was obtained. Figure~\ref{fig:im_sub} illustrates the PSF subtraction process.

\subsection{Photometric calibration and uncertainty}

As the stellar PSF peak is saturated for the main sequence of images, and since much image processing is done to subtract the stellar PSF from each image, special care must be taken to calibrate the photometry of the residual images and ensure that the contrast limits calculated are accurate.

When the PSF peak is saturated, relative photometry can be calibrated by scaling the stellar flux measured in the unsaturated images obtained before and/or after the saturated sequence according to the ratio of the exposure times of the saturated and unsaturated images. However, the accuracy of this calibration method is affected by the (unknown) variations in Strehl ratio, hence of the peak PSF flux, that may have occurred between the saturated and unsaturated observations. To mitigate this problem, the calibration approach we adopted relies on a sharp ghost artifact located $(+0.09\arcsec,-2.45\arcsec)$ from the PSF center in the ALTAIR/NIRI images. Since the intensity of this ghost artifact is proportional to the PSF intensity, it can be used to infer the peak flux of a saturated PSF. This was verified for all sequences for which both unsaturated and saturated data were available. First, the stellar flux was measured in the unsaturated images using a circular aperture of diameter equal to the FWHM of the PSF. When unsaturated data were acquired both before and after the saturated sequence, the mean of the two values was used. Then the flux of the ghost artifact in the same aperture was measured for each image of the saturated sequence. The median of these values, scaled according to the ratio of the exposure times of the saturated and unsaturated images, was then compared to the stellar flux, and the process was repeated for all sequences that include both saturated and unsaturated data. Similar values were found for all sequences; the mean ratio of the flux of the ghost over that of the PSF peak was found to be $6.1\times10^{-5}$, with a standard deviation of $0.6\times10^{-5}$. Comparisons of the flux of background stars bright enough to be visible in each individual image of a sequence with the flux of the ghost in the corresponding images also confirmed that the intensity of the ghost is indeed directly proportional to the intensity of off-axis sources.

The procedure used for calibrating the photometry was the following. The flux of the ghost was measured for each image of a sequence and the median of these values, divided by the ratio quoted above, was taken to represent the peak stellar PSF flux, ${F_\star}$. This calibration method should be more accurate than the one based solely on unsaturated data obtained before and/or after the saturated sequence because the median ghost flux is affected in the same way as the median of all residual images by the variations of Strehl ratio that may have occurred during the sequence of saturated images or between the saturated and unsaturated measurements. For this reason, this calibration was used even for the sequences for which unsaturated data were available.

Observations obtained with ALTAIR without the field lens suffer from important off-axis Strehl degradation because of anisoplanatism; this degradation must be taken into account when calculating contrast. Unfortunately, it is virtually impossible to quantify the specific degradation pertaining to our data as there are no bright reference off-axis point sources available for every sequence of images. Instead, we have used the average anisoplanetism Strehl ratio degradation formula indicated on the ALTAIR webpage\footnote{http://www.gemini.edu/sciops/instruments/altair/\\altairCommissioningPerformance.html}, which is $f_{\rm aniso}(\theta) \equiv S(\theta)/S_0=\mathrm{e}^{-(\theta/12.5)^2}$, where $S(\theta)$ is the Strehl ratio at angular separation $\theta$, expressed in arcseconds, and $S_0$ is the on-axis Strehl ratio. This factor was used to correct the noise and the flux of faint point sources measured in the residual images.

\begin{figure}
\epsscale{1}
\plotone{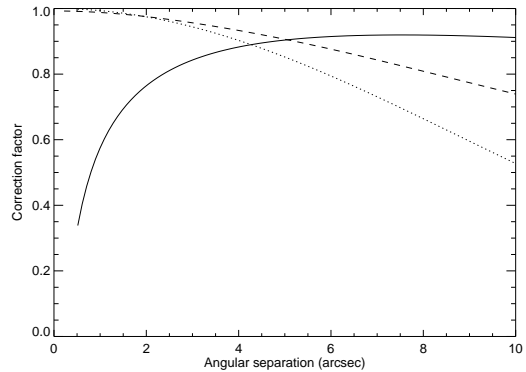}
\caption{\label{fig:cor} Typical values of $f_{\rm sub}$ ({\it solid line}), $f_{\rm aniso}$ ({\it dotted line}), and $f_{\rm sm}$ ({\it dashed line}) as a function of angular separation. The curves shown are for the target HD~166181.}
\end{figure}

As explained in \citet{lafreniere07}, while the subtraction of an optimized reference PSF obtained using the LOCI algorithm leads to better signal-to-noise (S/N) ratios, it removes partially the flux of the point sources sought after. This flux loss must be accounted for when calculating contrast. This is done by calculating the normalized residual intensity, $f_{\rm sub}$, of artificially implanted point sources after execution of the subtraction algorithm; the method used is described in \S 4.3 of \citet{lafreniere07}. Then using flux measurements made in the residual image, the factor $f_{\rm sub}$ is used to infer the true flux of a point source, i.e. that before execution of the subtraction algorithm.

Another effect that must be taken into account for ADI data is the azimuthal smearing of an off-axis point source that occurs as the field of view rotates during an integration; this causes a fraction of the source's flux to fall outside of the circular aperture used for photometric measurements. The amount of flux loss in the aperture was calculated for each \\
\begin{deluxetable}{lccccc}
\tablewidth{0pt}
\tablecolumns{6}
\tablecaption{Photometric uncertainties \label{tbl:sigphot}}
\tablehead{ \colhead{Sep. (\arcsec)} & \colhead{$<4$} & \colhead{$4-7$} & \colhead{$7-10$} & \colhead{$10-13$} & \colhead{$>13$} }
\startdata
$\sigma$ (mag) & 0.07 & 0.12 & 0.15 & 0.26 & 0.39
\enddata
\end{deluxetable}

\noindent sequence of images as follows. For a given angular separation and for each image of a sequence, a copy of the unsaturated PSF was smeared according to its displacement during an integration. When unsaturated data were unavailable, a 2D Gaussian of the appropriate FWHM was used in place of the unsaturated PSF. The median of these smeared PSFs was obtained and the flux in a circular aperture was measured. This flux was divided by the flux of the original PSF in the same aperture to obtain the smearing factor $f_{\rm sm}$, which is used to correct the flux or noise measured in the images.

Given all of these considerations, the contrast at angular separation $\theta$ was calculated as
\begin{equation}\label{eq:contrast}
C(\theta)=\frac{F(\theta)}{f_{\rm aniso}(\theta) f_{\rm sm}(\theta) f_{\rm sub}(\theta)} \times \frac{1}{F_\star},
\end{equation}
where $F(\theta)$ is either the noise or the flux of a point source in a circular aperture of diameter equal to one PSF FWHM, at angular separation $\theta$, in the residual image. Note that the contrast in the equation above is defined such that a fainter companion, or a smaller residual noise, has a smaller contrast value. Eq.~(\ref{eq:contrast}) was used for all contrast calculations in the present work. Typical correction factors as a function of angular separation are shown in Figure~\ref{fig:cor}.

An estimate of the photometric accuracy resulting from the entire process was obtained by calculating the mean absolute difference between the magnitudes calculated at two epochs for every faint background star that was observed twice (see \S\ref{sect:cand}); this mean absolute difference was taken to represent $\sqrt{2}$ times the photometric uncertainty. This photometric uncertainty was found to vary significantly with angular separation, indicating that it is dominated by the uncertainty on the anisoplanatism factor. The photometric uncertainty as a function of angular separation is reported in Table~\ref{tbl:sigphot}; it is typically 0.07--0.15~mag for separations below 10\arcsec. For completeness, it is noted that a higher photometric uncertainty, by about 0.08~mag, results when the unsaturated data obtained before and/or after the main sequence of saturated images are used to determine $F_\star$, rather than the median flux of the ghost artifact, justifying our choice to use the calibration based on the flux of the ghost for all sequences.

\section{Results}\label{sect:results}

\subsection{Detection limits}\label{sect:limits}

Detection limits are based on a measure of the noise in the residual images. To calculate this noise, the residual images were first convolved by a circular aperture of diameter equal to one PSF FWHM, which is typically $\sim$0.07\arcsec, and the noise as a function of angular separation from the image center, $F(\theta)$, was determined as the standard deviation of the pixel values in an annulus of width equal to one PSF FWHM. As shown in \citet{lafreniere07} and \citet{marois07b}, the noise in an ADI residual image has a distribution similar to a Gaussian; using a $5\sigma$ detection threshold is thus appropriate for our data to limit the number of false positives. Given that a residual image typically contains $\sim$2$\times10^5$ resolution elements, roughly 0.1 false positive per target is expected on average. Because of the underlying noise in a residual image, some sources near the detection threshold might not be detected. From Gaussian statistics, the probability that the residual signal underlying a source is below $0\sigma$, $-1\sigma$, or $-2\sigma$ is 50\%, 16\%, or 2.3\%, respectively. Our detection completeness for sources whose true intensities are 5$\sigma$, 6$\sigma$, or $>$7$\sigma$ is thus 50\%, 84\%, or $>$97.7\%, respectively. Note that for a similar reason, some sources whose true intensities are below the 5$\sigma$ threshold could be detected as well. These effects will be taken into account appropriately in the statistical analysis of the results presented in \S\ref{sect:discussion}.

\begin{figure*}
\epsscale{1}
\plotone{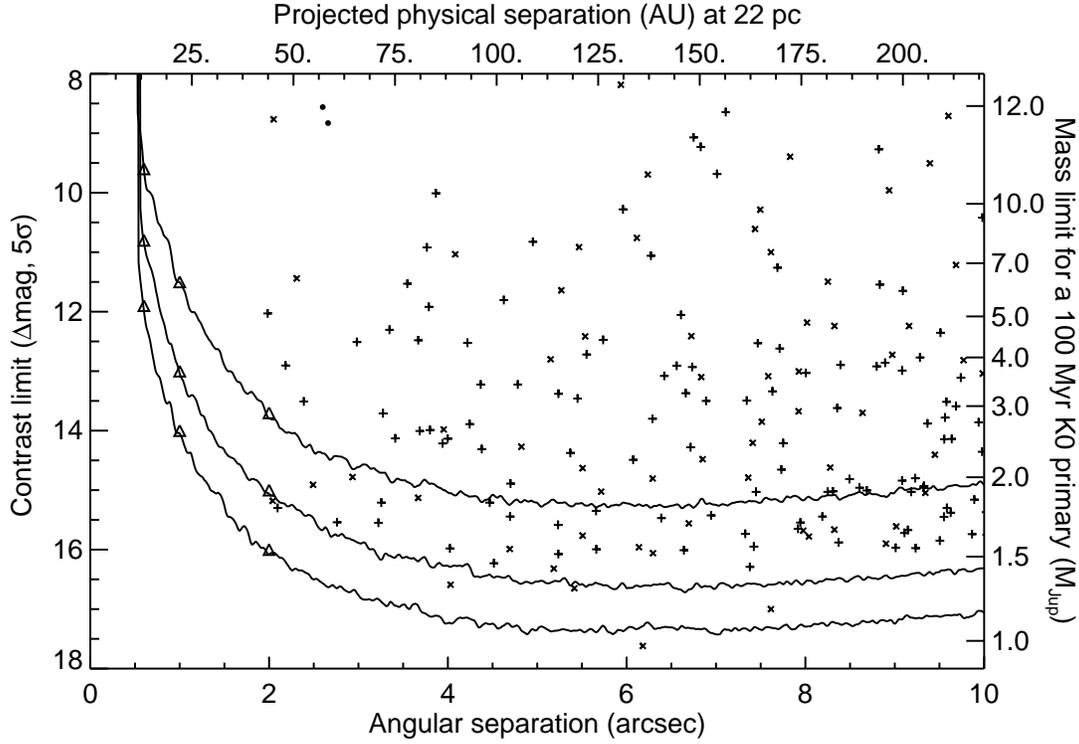}
\caption{\label{fig:limits} Survey detection limits in difference of magnitude (in the NIRI CH4-short filter) between an off-axis point source and the target star, at the $6\sigma$ level. The top, middle, and bottom curves are respectively for the targets GJ~507.1, HD~166181, and HD~208313, which are representative of poor, median, and good performance reached by the survey. Companion candidates identified around targets of galactic latitude $|b|<15$ are shown by $+$ symbols, while those identified around targets with $|b|\ge15$ are shown by $\times$ symbols. The two filled circles near (2.6,8.6) indicate the components of the binary brown dwarf companion to HD~130948. The fiducial point sources shown in Fig.~\ref{fig:fakesources} are marked with triangles. The top and right axes show, for reference only, the projected separation in AU and the detection limits in \mjup\ that would apply for a 100 Myr old K0 star located 22~pc away.}
\end{figure*}

The detection limits achieved for all target stars, expressed in magnitude difference, are presented in Table~\ref{tbl:limits}. The last two lines of this table present the median and best contrast, over the 85 observations, achieved at each angular separation. The median detection limits in magnitude difference are 9.5 at 0.5\arcsec, 12.9 at 1\arcsec, 15.0 at 2\arcsec, and 16.5 at 5\arcsec. The detection limits are presented graphically in Figure~\ref{fig:limits} for the stars HD~208313, HD~166181, and GJ~507.1, which are representative of poor, median, and good contrast performance, respectively.

\begin{figure*}
\epsscale{1}
\plotone{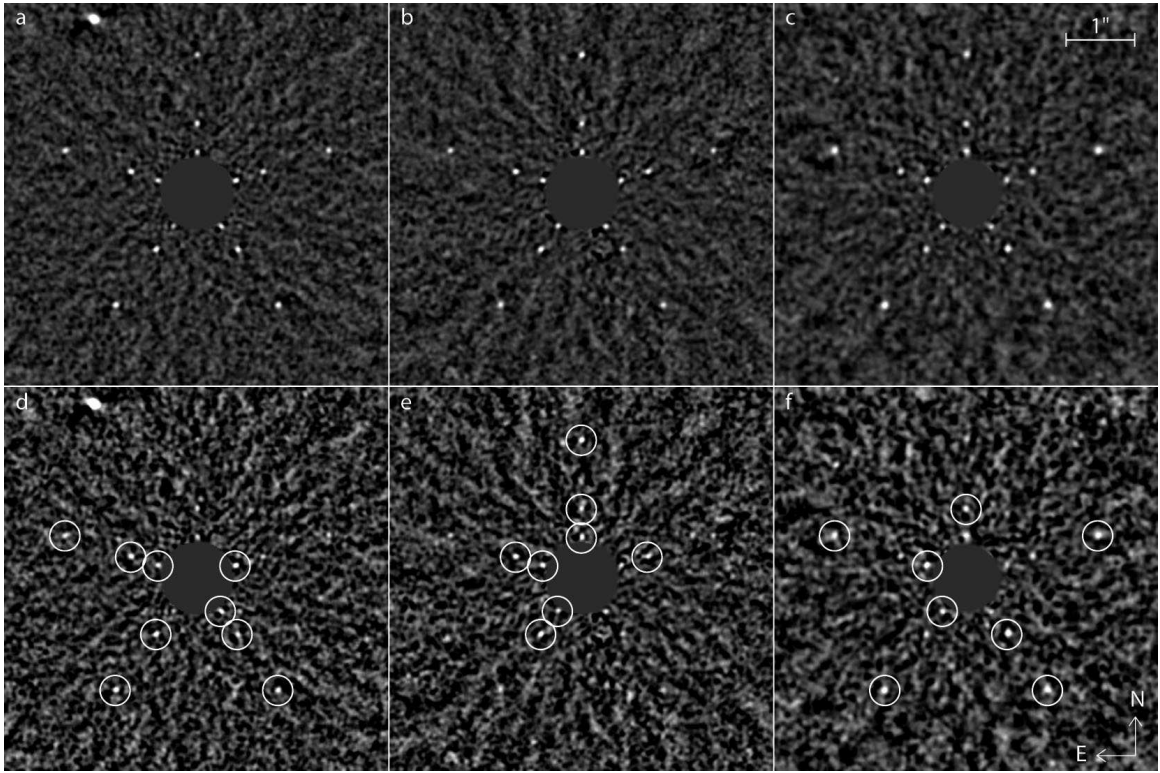}
\caption{\label{fig:fakesources} Final S/N residual images for three sequences of images to which fiducial point sources have been implanted. The fiducial point sources have been added at 5 P.A.'s (0\degr, 72\degr, 144\degr, 216\degr, and 288\degr) and 3 angular separations (0.6\arcsec, 1.0\arcsec, and 2.0\arcsec). For bottom panels (d--f) the intensity of each source was set to the corresponding detection limit (5$\sigma$) indicated in Table~\ref{tbl:limits}, while it was set 0.75 mag brighter (i.e.~10$\sigma$) for top panels (a--c). Panels (a,d), (b,e), and (c,f) are for the stars HD~208313, HD~166181, and GJ~507.1, respectively. The bright spot at the upper left corner of panels (a,d) is a real background star. The display intensity scale is linear from $-2$ to $+10$ for top panels (a--c), and from $-1$ to $+5$ for bottom panels (d--f). In bottom panels (d--f), the sources that would have been detected (S/N$\ge$5) have been circled in white. According to expectations, the detection completeness is roughly 50\% for sources whose true intensity is equal to the 5$\sigma$ detection limit.}
\end{figure*}

For consistency we have verified the validity of these detection limits by implanting fiducial sources in the sequence of original images and then processing the data as described in \S\ref{sect:datared}. An example, incorporating artificial sources at the 5$\sigma$ and 10$\sigma$ levels at various separations, is shown in Figure~\ref{fig:fakesources} for the stars HD~208313, HD~166181, and GJ~507.1. As visible in this figure, sources exactly at our detection limits can indeed be detected with the expected completeness level.

One must resort to evolution models of giant planets to convert the detection limits mentioned above into mass limits. Traditionally, such evolution models have assumed arbitrary initial conditions for the planets \citep[e.g.][]{baraffe03,burrows97}, with the caution that their results depend on the specific initial conditions adopted for ages below a few million years \citep{baraffe02}. Recent evolution models \citep{marley07} that incorporate initial conditions calculated explicitly for planets formed through core accretion indicate that it may in fact take as much as 10--100~Myr before the planets ``forget'' their initial conditions; the effect being more important for more massive planets. Nevertheless, given the typical ages of our target stars (50--300 Myr) and the good contrast limits we have reached, the different evolution models should yield similar mass detection limit estimates. As a simple example, consider a contrast of 12.9~mag in the NIRI CH4-short filter around a K0 star (typical at a separation of 1\arcsec). The ``hot start'' models of \citet{baraffe03} would give masses of 2.6~\mjup\ and 3.9~\mjup\ at 50~Myr and 100~Myr, respectively, while the ``core accretion'' models of \citet{marley07} would give masses of $\sim$3.0~\mjup\ and $\sim$4.5~\mjup, respectively\footnote{For this simple calculation, it was assumed that the luminosity ratios between the ``hot start'' and ``core accretion'' models were representative of the $H$-band magnitude differences.}. The difference between the models would be smaller for smaller masses (better contrast limits (i.e. beyond $\sim$1\arcsec) and/or greater ages), while it would be larger for larger masses (worse contrast limits and/or smaller ages). In this work, keeping the latter caveat in mind, we have used the COND evolution models of \citet{baraffe03}, for which absolute $H$-band magnitudes as a function of mass and age are readily available. The following procedure was used to estimate the contrast, in the \\

\onecolumngrid
\begin{deluxetable}{lccccccccccccc}
\tablewidth{200pt}
\tabletypesize{\scriptsize}
\tablecolumns{14}
\tablecaption{GDPS detection limits\tablenotemark{a} \label{tbl:limits}}
\tablehead{
\colhead{Name} & \colhead{0.50\arcsec} & \colhead{0.60\arcsec} & \colhead{0.75\arcsec} & \colhead{1.00\arcsec} & \colhead{1.25\arcsec} & \colhead{1.50\arcsec} & \colhead{2.00\arcsec} & \colhead{2.50\arcsec} & \colhead{3.00\arcsec} & \colhead{4.00\arcsec} & \colhead{5.00\arcsec} & \colhead{7.50\arcsec} & \colhead{10.00\arcsec}
}
\startdata
HD 166 & - & - & - & 12.5 & 13.1 & 13.9 & 14.9 & 15.4 & 15.9 & 16.5 & 16.9 & 17.3 & 17.3 \\
HD 691 & - & 11.1 & 12.1 & 13.2 & 14.1 & 14.7 & 15.6 & 15.9 & 16.2 & 16.6 & 16.7 & 16.6 & 16.3 \\
HD 1405 &  9.2 & 10.5 & 11.4 & 12.7 & 13.5 & 14.0 & 14.8 & 15.3 & 15.7 & 16.0 & 16.1 & 16.1 & 15.8 \\
HD 5996 & - & 10.8 & 12.0 & 13.2 & 14.1 & 14.6 & 15.4 & 15.8 & 16.1 & 16.5 & 16.6 & 16.6 & 16.4 \\
HD 9540 & - & - & 11.8 & 13.1 & 14.0 & 14.5 & 15.4 & 16.0 & 16.4 & 17.0 & 17.3 & 17.6 & 17.5 \\
HD 10008 & - & 10.0 & 11.2 & 12.4 & 13.2 & 13.8 & 14.7 & 15.2 & 15.6 & 16.2 & 16.5 & 16.5 & 16.3 \\
GJ 82 &  8.9 &  9.5 & 10.5 & 11.8 & 12.5 & 13.2 & 13.7 & 14.3 & 14.6 & 14.9 & 15.0 & 14.8 & 14.6 \\
HD 14802 & - & - & - & - & 11.8 & 12.4 & 13.3 & 14.0 & 14.7 & 15.8 & 16.8 & 17.4 & 17.9 \\
HD 16765 & - & - & - & 13.0 & 13.9 & 14.5 & 15.3 & 15.8 & 16.2 & 16.9 & 17.4 & 17.5 & 17.6 \\
HD 17190 & - & 10.5 & 12.2 & 13.7 & 14.2 & 14.8 & 15.5 & 15.9 & 16.3 & 16.6 & 16.8 & 16.6 & 16.2 \\
HD 17382 & - & 10.8 & 12.0 & 13.3 & 14.1 & 14.6 & 15.4 & 15.9 & 16.3 & 16.8 & 17.0 & 17.0 & 16.7 \\
HD 17925 & - & - & 11.9 & 13.6 & 14.6 & 15.4 & 16.2 & 16.8 & 17.1 & 17.6 & 17.7 & 17.7 & 17.4 \\
HD 18803 & - & - & 11.3 & 12.9 & 13.8 & 14.5 & 15.5 & 16.0 & 16.5 & 16.8 & 17.1 & 17.2 & 16.9 \\
HD 19994 & - & - & - & 13.5 & 14.3 & 15.0 & 15.8 & 16.4 & 16.7 & 17.4 & 17.8 & 18.3 & 18.4 \\
HD 20367 & - & - & - & 11.6 & 12.2 & 12.8 & 13.9 & 14.4 & 14.8 & 15.6 & 16.0 & 16.3 & 16.1 \\
2E 759 &  8.6 &  9.4 &  9.9 & 11.0 & 11.8 & 12.2 & 13.0 & 13.4 & 13.6 & 13.9 & 14.0 & 13.9 & 13.6 \\
HD 22049 & - & - & - & - & - & - & - & 15.9 & 16.5 & 17.3 & 17.7 & 18.5 & 18.9 \\
HIP 17695 & 10.0 & 10.8 & 11.8 & 12.8 & 13.6 & 14.2 & 14.8 & 15.1 & 15.3 & 15.7 & 15.7 & 15.6 & 15.3 \\
HD 25457 & - & - & - & - & 12.5 & 13.1 & 13.9 & 14.8 & 15.2 & 16.0 & 16.6 & 17.0 & 17.0 \\
HD 283750 & - & - & 12.2 & 13.4 & 14.2 & 15.1 & 15.9 & 16.4 & 16.8 & 17.2 & 17.2 & 17.1 & 16.7 \\
HD 30652 & - & - & - & - & - & - & 14.9 & 15.5 & 15.9 & 16.7 & 17.3 & 18.2 & 18.6 \\
GJ 182 & 10.0 & 10.5 & 11.9 & 13.1 & 14.0 & 14.7 & 15.4 & 15.8 & 16.1 & 16.4 & 16.5 & 16.4 & 16.2 \\
GJ 234A &  9.5 & 10.1 & 11.2 & 12.3 & 13.3 & 13.9 & 14.6 & 15.1 & 15.4 & 15.9 & 16.2 & 16.3 & 16.1 \\
GJ 281 & - &  9.0 & 10.4 & 12.0 & 12.9 & 13.5 & 14.3 & 14.6 & 15.0 & 15.3 & 15.3 & 15.4 & 15.2 \\
GJ 285 & - &  8.0 & 10.1 & 11.6 & 12.6 & 13.3 & 13.8 & 14.5 & 14.9 & 15.5 & 15.8 & 15.9 & 15.8 \\
HD 72905 & - & - & - & 11.2 & 12.5 & 13.1 & 14.2 & 14.9 & 15.4 & 16.3 & 16.7 & 17.4 & 17.7 \\
HD 75332 & - & - & 10.8 & 12.3 & 13.0 & 13.9 & 14.9 & 15.5 & 15.7 & 16.6 & 17.1 & 17.4 & 17.3 \\
HD 77407 & - & - & 10.3 & 11.4 & 12.3 & 13.0 & 14.0 & 14.8 & 15.0 & 15.7 & 16.0 & 16.3 & 16.2 \\
HD 78141 & - & - & 11.5 & 13.0 & 13.7 & 14.5 & 15.4 & 15.8 & 16.1 & 16.5 & 16.6 & 16.5 & 16.3 \\
HD 82558 & - & - & 11.5 & 12.9 & 13.8 & 14.4 & 15.4 & 15.9 & 16.1 & 16.6 & 16.8 & 17.0 & 16.7 \\
HD 82443 & - & - & 11.5 & 13.0 & 14.1 & 14.8 & 15.9 & 16.4 & 16.8 & 17.2 & 17.5 & 17.7 & 17.5 \\
GJ 393 & - & - & 11.8 & 13.3 & 14.1 & 14.6 & 15.6 & 16.0 & 16.2 & 16.7 & 16.8 & 16.9 & 16.8 \\
HD 90905 & - & - & 11.4 & 12.7 & 13.7 & 14.1 & 15.1 & 15.7 & 16.1 & 16.5 & 16.6 & 16.6 & 16.4 \\
HD 91901 & - &  9.2 & 10.0 & 11.4 & 12.1 & 12.8 & 13.6 & 14.1 & 14.4 & 14.9 & 14.8 & 14.8 & 14.6 \\
HD 92945 & - & - & 10.8 & 12.1 & 13.0 & 13.8 & 14.6 & 15.1 & 15.5 & 15.9 & 16.1 & 16.3 & 16.1 \\
HD 93528 &  8.5 &  9.3 & 10.2 & 11.6 & 12.6 & 13.3 & 14.2 & 14.8 & 15.0 & 15.5 & 15.7 & 15.9 & 15.7 \\
GJ 402 &  8.4 &  9.2 & 10.5 & 11.6 & 12.5 & 13.1 & 14.0 & 14.5 & 14.9 & 15.4 & 15.4 & 15.6 & 15.3 \\
HD 96064 & - & 10.9 & 12.3 & 13.5 & 14.3 & 14.9 & 15.6 & 16.1 & 16.3 & 16.6 & 16.8 & 16.8 & 16.6 \\
HD 97334 & - & - & - & 13.6 & 14.7 & 15.1 & 16.0 & 16.4 & 16.7 & 17.2 & 17.4 & 17.5 & 17.3 \\
HD 102195 &  9.8 & 11.2 & 12.2 & 13.3 & 14.1 & 14.7 & 15.4 & 15.9 & 16.1 & 16.5 & 16.6 & 16.6 & 16.3 \\
HD 102392 &  9.5 & 10.3 & 11.4 & 12.6 & 13.5 & 13.9 & 14.7 & 15.3 & 15.6 & 16.1 & 16.3 & 16.3 & 16.2 \\
HD 105631 & - & - & 11.7 & 12.8 & 13.5 & 14.2 & 15.1 & 15.5 & 16.0 & 16.4 & 16.7 & 16.8 & 16.5 \\
HD 107146 & - & - & 11.7 & 12.5 & 13.5 & 14.0 & 15.0 & 15.4 & 15.8 & 16.2 & 16.5 & 16.5 & 16.3 \\
HD 108767B &  8.4 &  9.7 & 10.6 & 11.9 & 12.8 & 13.5 & 14.3 & 14.9 & 15.1 & 15.7 & 15.8 & 16.0 & 15.7 \\
HD 109085 & - & - & - & - & 13.4 & 14.0 & 14.9 & 15.8 & 16.3 & 17.2 & 17.7 & 18.3 & 18.5 \\
BD+60 1417 & 10.0 & 11.1 & 12.0 & 13.0 & 13.8 & 14.2 & 14.7 & 15.0 & 15.3 & 15.5 & 15.5 & 15.4 & 15.1 \\
HD 111395 & - & - & - & 13.4 & 14.3 & 15.0 & 15.9 & 16.4 & 16.7 & 17.2 & 17.4 & 17.6 & 17.3 \\
HD 113449 & - & - & 11.5 & 12.6 & 13.7 & 13.9 & 14.9 & 15.4 & 15.7 & 16.3 & 16.5 & 16.6 & 16.4 \\
GJ 507.1 & - &  9.6 & 10.4 & 11.5 & 12.2 & 12.9 & 13.7 & 14.3 & 14.6 & 15.0 & 15.2 & 15.2 & 14.9 \\
HD 116956 & - & - & 11.3 & 12.7 & 13.5 & 14.2 & 15.1 & 15.7 & 16.0 & 16.5 & 16.7 & 16.8 & 16.6 \\
HD 118100 &  8.4 &  9.4 & 10.5 & 11.6 & 12.3 & 12.8 & 13.5 & 14.0 & 14.1 & 14.4 & 14.5 & 14.4 & 14.2 \\
GJ 524.1 & 10.1 & 11.0 & 12.0 & 13.0 & 13.6 & 14.2 & 14.9 & 15.2 & 15.4 & 15.4 & 15.5 & 15.4 & 15.0 \\
HD 124106 & - & 10.3 & 11.6 & 13.0 & 13.8 & 14.4 & 15.4 & 15.7 & 15.9 & 16.4 & 16.7 & 16.8 & 16.6 \\
HD 125161B & 10.5 & 11.3 & 12.4 & 13.6 & 14.3 & 14.6 & 15.4 & 15.8 & 16.0 & 16.2 & 16.4 & 16.3 & 16.1 \\
HD 129333 & - & 10.7 & 11.7 & 13.2 & 13.9 & 14.4 & 15.3 & 15.7 & 16.2 & 16.4 & 16.7 & 16.7 & 16.5 \\
HD 130004 & - & - & 12.0 & 13.1 & 14.1 & 14.5 & 15.3 & 15.8 & 16.1 & 16.5 & 16.7 & 16.7 & 16.4 \\
HD 130322 & - & 11.1 & 12.1 & 13.2 & 13.9 & 14.3 & 15.2 & 15.6 & 15.9 & 16.3 & 16.4 & 16.4 & 16.2 \\
HD 130948 & - & - & - & 12.4 & 13.2 & 13.8 & 14.7 & 15.4 & 15.7 & 16.5 & 16.9 & 17.3 & 17.3 \\
HD 135363 & - &  9.2 & 10.9 & 12.3 & 13.1 & 13.7 & 14.6 & 15.1 & 15.3 & 15.5 & 15.7 & 15.6 & 15.4 \\
HD 139813 & - & - & 10.3 & 11.2 & 11.9 & 12.6 & 13.6 & 14.3 & 14.9 & 15.7 & 16.1 & 16.3 & 16.1 \\
HD 141272 & - & - & 12.2 & 13.7 & 14.4 & 15.0 & 15.8 & 16.2 & 16.5 & 16.9 & 16.9 & 17.1 & 16.9 \\
HD 147379B & - & 10.0 & 11.3 & 12.8 & 13.5 & 14.1 & 15.0 & 15.3 & 15.6 & 15.8 & 16.0 & 16.0 & 15.7 \\
GJ 628 & - & - & 10.4 & 12.2 & 13.0 & 13.7 & 14.6 & 15.2 & 15.7 & 16.2 & 16.6 & 16.9 & 16.7 \\
HIP 81084 &  9.5 & 10.3 & 11.4 & 12.3 & 13.0 & 13.5 & 14.0 & 14.4 & 14.6 & 14.7 & 14.7 & 14.6 & 14.3 \\
HD 160934 &  9.5 & 10.1 & 11.2 & 12.5 & 13.3 & 13.9 & 14.6 & 14.9 & 15.0 & 15.3 & 15.3 & 15.2 & 14.9 \\
HD 162283 & 10.3 & 11.2 & 12.2 & 13.4 & 14.0 & 14.6 & 15.2 & 15.7 & 16.1 & 16.4 & 16.5 & 16.5 & 16.1 \\
HD 166181 & - & 10.8 & 11.7 & 13.0 & 13.7 & 14.3 & 15.0 & 15.4 & 15.8 & 16.2 & 16.5 & 16.5 & 16.3 \\
HD 167605 &  9.4 & 10.5 & 11.4 & 12.5 & 13.3 & 14.0 & 14.8 & 15.1 & 15.6 & 15.9 & 16.1 & 16.1 & 15.9 \\
HD 187748 & - & 10.8 & 11.7 & 12.9 & 13.7 & 14.5 & 15.3 & 15.9 & 16.3 & 17.0 & 17.3 & 17.6 & 17.4 \\
GJ 791.3 &  9.6 & 11.0 & 12.0 & 13.3 & 13.8 & 14.4 & 15.1 & 15.6 & 15.7 & 16.0 & 16.1 & 16.1 & 15.7 \\
HD 197481 & - & - & - & 11.0 & 11.7 & 12.4 & 13.5 & 14.3 & 14.7 & 15.5 & 16.1 & 16.4 & 16.3 \\
HD 201651 & 10.1 & 11.4 & 12.3 & 13.3 & 14.1 & 14.6 & 15.3 & 15.8 & 16.1 & 16.4 & 16.5 & 16.5 & 16.3 \\
HD 202575 & - & - & 11.4 & 12.5 & 13.3 & 14.0 & 14.9 & 15.5 & 16.1 & 16.6 & 16.8 & 17.0 & 16.7 \\
GJ 4199 & 10.5 & 11.2 & 12.0 & 13.2 & 13.8 & 14.5 & 15.1 & 15.6 & 15.7 & 16.0 & 16.1 & 15.8 & 15.4 \\
HD 206860 & - & - & 12.2 & 13.3 & 13.8 & 14.5 & 15.2 & 15.7 & 16.0 & 16.5 & 16.9 & 17.0 & 16.7 \\
HD 208313 & - & 11.9 & 13.0 & 14.0 & 14.7 & 15.2 & 16.0 & 16.5 & 16.7 & 17.2 & 17.3 & 17.3 & 17.0 \\
V383 Lac & 10.2 & 11.0 & 11.9 & 13.0 & 13.6 & 14.3 & 14.8 & 15.2 & 15.5 & 15.9 & 16.1 & 16.0 & 15.8 \\
HD 213845 & - & - & - & 13.3 & 14.0 & 14.7 & 15.7 & 16.3 & 16.8 & 17.2 & 17.6 & 18.1 & 18.0 \\
GJ 875.1 &  9.6 & 10.5 & 11.2 & 12.3 & 13.1 & 13.5 & 14.4 & 14.9 & 15.1 & 15.5 & 15.6 & 15.5 & 15.1 \\
GJ 876 & - & - & - & 11.0 & 12.2 & 12.6 & 13.7 & 14.3 & 15.1 & 15.8 & 16.2 & 16.6 & 16.6 \\
GJ 9809 & 11.3 & 12.1 & 12.8 & 14.0 & 14.6 & 15.0 & 15.5 & 15.9 & 15.9 & 16.2 & 16.3 & 16.1 & 15.8 \\
HD 220140 & - & - & 12.0 & 13.1 & 13.9 & 14.5 & 15.3 & 15.8 & 16.1 & 16.4 & 16.6 & 16.5 & 16.3 \\
HD 221503 & - & 10.4 & 11.8 & 13.2 & 14.1 & 14.6 & 15.3 & 15.8 & 16.2 & 16.7 & 17.0 & 17.1 & 17.0 \\
GJ 900 &  8.9 & 10.1 & 10.8 & 12.4 & 13.2 & 13.9 & 14.9 & 15.4 & 15.8 & 16.1 & 16.2 & 16.1 & 16.0 \\
GJ 907.1 &  8.4 &  9.0 & 10.0 & 11.2 & 12.1 & 12.5 & 13.4 & 14.0 & 14.3 & 14.8 & 15.0 & 15.1 & 14.9 \\
\hline
Median &  9.5 & 10.5 & 11.5 & 12.9 & 13.6 & 14.2 & 15.0 & 15.5 & 15.8 & 16.3 & 16.5 & 16.5 & 16.3 \\
\hline
Best & 11.3 & 12.1 & 13.0 & 14.0 & 14.7 & 15.4 & 16.2 & 16.8 & 17.1 & 17.6 & 17.8 & 18.5 & 18.9
\enddata
\tablenotetext{a}{Magnitude difference in the NIRI CH4-short filter, at a 5$\sigma$ level.}
\end{deluxetable}
\twocolumngrid

\noindent NIRI CH4-short filter, of a planet of given mass orbiting a given target.  The absolute $H$-band magnitude of the planet was first obtained directly from the evolution models of \citet{baraffe03} and converted into an apparent magnitude, $H_{\rm pl}$, using the distance of the star. The corresponding magnitude in the NIRI CH4-short filter was then calculated as
\begin{equation}
m_{\rm pl} = H_{\rm pl} - 2.5 \log \left( \frac{f_{\rm CH4}}{f_H} \right),
\end{equation}
where $f_{\rm CH4}$ and $f_H$ are the mean flux density of the planet in the NIRI CH4-short and broad band $H$ filters, respectively; their values were calculated using a synthetic spectrum of appropriate effective temperature and surface gravity \citep{baraffe03,allard01}\footnote{Spectra available at ftp://ftp.ens-lyon.fr/pub/users/CRAL/fallard/}. We recall here (c.f.~\S\ref{sect:obs}) that the ratio $\frac{f_{\rm CH4}}{f_H}$ is typically 1.5--2.5 for giant planets depending on their mass and age.
The stellar magnitudes in the NIRI CH4-short and broad band $H$ filters were assumed to be equal, such that the contrast of the planet was obtained as $m_{\rm pl}-H_\star$, where $H_\star$ is the $H$-band apparent magnitude of the target star. The 5$\sigma$ contrast levels of planets of various masses orbiting a K0 primary of 100~Myr,\footnote{An $H$-band absolute magnitude of 4.0 was used, this is the mean value of the K0 stars in the sample.} the typical target of the survey, are presented in Figure~\ref{fig:limits}. For a typical target located at 22~pc from the Sun, the median detection limits correspond to 10.8~\mjup\ at 11~AU, 3.9~\mjup\ at 22~AU, 1.9~\mjup\ at 44~AU, and 1.4~\mjup\ at 110~AU.

The typical contrast reached by our survey improves on earlier surveys \citep[e.g.][]{lowrance05, masciadri05, chauvin06, biller07} by at least 1 mag at 1\arcsec, 1.5 mag at 2\arcsec, and $\sim3$ mag at larger separations. For the 27 targets for which our data were in the linear regime of the detector at a separation of 0.5\arcsec, our detection limits at this separation are similar to those achieved with the SDI device at the Very Large Telescope \citep{biller07}. The contrast reached by GDPS observations is the highest that has been achieved to date at separations larger than $\sim$0.7\arcsec.

\subsection{Candidate companion detections}\label{sect:cand}

To identify candidate companions, the residual images were first convolved by a circular aperture of diameter equal to one PSF FWHM, and then converted to signal-to-noise (S/N) images that were visually inspected for point sources at a $\gtrsim$5$\sigma$ level. After identification of a point source, its position was measured by fitting a 2D Gaussian function, and its flux was measured in an aperture of diameter equal to one PSF FWHM; both operations were done in the non-convolved residual image. The contrast of the point source was then calculated using Eq.(\ref{eq:contrast}). More than 300 faint point sources were found around 54 targets, 188 of which are found around only 7 stars located at low galactic latitudes ($b < 11\degr$). Up to now, all but six of the 54 stars with candidates were re-observed at a subsequent epoch to verify whether or not the faint point sources detected are co-moving with the target star.

\begin{figure}
\epsscale{1}
\plotone{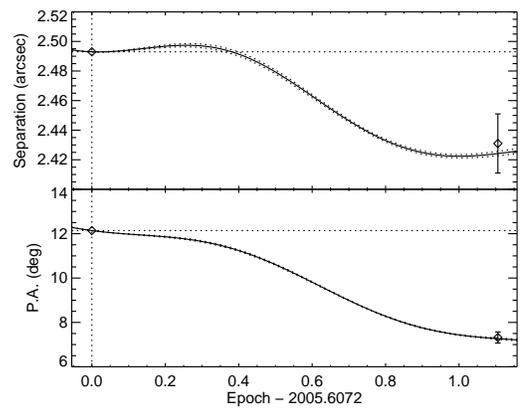}
\caption{\label{fig:pm} Verification of the background nature of the point source detected around the young star HD~691. Open diamonds mark the observed separation ({\it top}) and P.A. ({\it bottom}) of the point source at the two epochs. The solid line indicates the expected separation and P.A. of a distant background source as a function of time. The observations agree very well with the expected motion of a background source, indicating that the source is not associated with HD~691.}
\end{figure}

All candidate exoplanets observed at two epochs have been confirmed to be background sources by comparing their displacement between the two epochs with the expected displacement of a distant background source, based on the proper motion and parallax of the target; an example of this verification is presented in Figure~\ref{fig:pm}. As a reference for future planet searches, a compilation of all faint point sources identified around our target stars is presented in Table~\ref{tbl:cand}.

An estimate of the uncertainties on the measured separations and P.A. was obtained by calculating the mean absolute difference between the separation and P.A. measured at the second epoch and those predicted for this epoch based on the parallax and proper motion of the target stars. Given the high precision on the parallax and proper motion of the target stars, the differences observed are dominated by our measurement uncertainties. The mean absolute differences calculated are taken to represent $\sqrt{2}$ times the uncertainties; values of $\sigma_{\rm sep}=0.015\arcsec$ and $\sigma_{\rm P.A.}=0.2\degr$ are found.

\onecolumngrid
\begin{deluxetable}{lcccc}
\tablewidth{0pt}
\tablecolumns{5}
\tablecaption{Point sources detected \label{tbl:cand}}
\tablehead{
\colhead{Star} & \colhead{Epoch} & \colhead{Separation\tablenotemark{a}} & \colhead{P.A.\tablenotemark{b}} & \colhead{$\Delta m$\tablenotemark{c}} \\
\colhead{} & \colhead{} & \colhead{(arcsec)} & \colhead{(deg)} & \colhead{mag}
}
\startdata
    HD 166 & 2005.6482 & 10.23 &  82.9 & 12.60 \\
    HD 691 & 2005.6072 &  2.49 &  12.1 & 14.91 \\
   HD 1405 & 2004.6409 &  3.95 & 254.0 & 13.98 \\
   HD 5996 & 2005.6128 &  2.98 & 118.6 & 12.51 \\
           & 2005.6128 &  4.78 &  71.6 & 13.23 \\
           & 2005.6128 &  5.66 & 268.9 & 15.99 \\
           & 2005.6128 &  6.95 &  73.9 & 15.43 \\
           & 2005.6128 &  9.11 & 280.8 & 15.72\tablenotemark{e} \\
           & 2005.6128 &  9.15 & 228.1 & 15.67 \\
           & 2005.6128 &  9.50 & 120.6 & 15.85\tablenotemark{e} \\
           & 2005.6128 &  9.58 & 205.2 & 13.51 \\
           & 2005.6128 &  9.94 & 355.6 & 13.86 \\
           & 2005.6128 & 10.33 & 221.6 & 10.86 \\
           & 2005.6128 & 10.49 & 320.2 & 10.93 \\
           & 2005.6128 & 10.55 & 296.5 & 14.41 \\
           & 2005.6128 & 11.18 & 218.9 & 13.51 \\
           & 2005.6128 & 13.09 & 215.7 & 14.63 \\
           & 2005.6128 & 14.05 & 141.1 & 13.35 \\
           & 2005.6128 & 14.57 & 314.9 & 11.48 \\
           & 2005.6128 & 15.06 & 137.5 & 11.48 \\
   HD 9540 & 2005.6182 &  5.51 & 308.8 & 14.63 \\
           & 2005.6182 &  6.70 & 120.9 & 15.56 \\
     GJ 82 & 2005.6647 &  4.24 &  78.2 & 13.89\tablenotemark{d} \\
           & 2005.6647 &  5.45 &  35.3 & 13.46\tablenotemark{d} \\
           & 2005.6647 &  6.27 & 157.0 & 11.06\tablenotemark{d} \\
           & 2005.6647 &  6.29 & 228.1 & 13.80\tablenotemark{d} \\
           & 2005.6647 &  6.42 & 307.0 & 13.08\tablenotemark{d} \\
           & 2005.6647 &  6.75 & 105.7 &  9.07\tablenotemark{d} \\
           & 2005.6647 &  6.83 & 106.4 &  9.23\tablenotemark{d} \\
           & 2005.6647 &  6.95 &  25.7 &  6.57\tablenotemark{d} \\
           & 2005.6647 &  7.63 & 117.2 & 13.34\tablenotemark{d} \\
           & 2005.6647 &  8.82 & 315.9 &  9.27\tablenotemark{d} \\
           & 2005.6647 &  9.68 & 318.7 & 13.59\tablenotemark{d} \\
           & 2005.6647 &  9.74 & 334.0 & 13.11\tablenotemark{d} \\
           & 2005.6647 & 11.37 & 228.3 & 12.52\tablenotemark{d} \\
  HD 17382 & 2004.9740 & 11.78 & 130.8 & 13.16 \\
  HD 18803 & 2004.9795 &  7.61 & 166.1 & 17.00 \\
           & 2004.9795 &  7.98 & 208.3 & 15.68 \\
           & 2004.9795 & 10.36 &  52.8 & 15.10 \\
  HD 19994 & 2005.6648 &  6.18 & 187.4 & 17.62 \\
           & 2005.6648 &  6.30 & 185.3 & 16.06 \\
           & 2005.6648 & 11.64 &  72.7 & 17.51 \\
 HD 283750 & 2004.8132 &  7.73 & 175.8 & 14.65 \\
           & 2004.8132 & 12.72 & 104.2 & 13.85 \\
  HD 30652 & 2005.6978 &  2.04 & 106.3 & 15.18\tablenotemark{d} \\
           & 2005.6978 &  9.53 & 241.4 & 18.33\tablenotemark{d} \\
    GJ 182 & 2004.8459 &  5.15 & 220.3 & 12.80 \\
           & 2004.8459 &  7.44 & 233.7 & 10.61 \\
   GJ 234A & 2005.8455 &  3.27 &  48.8 & 13.71\tablenotemark{d} \\
           & 2005.8455 &  6.64 & 304.9 & 16.01\tablenotemark{d} \\
           & 2005.8455 &  7.45 & 215.1 & 15.03\tablenotemark{d} \\
           & 2005.8455 & 10.08 & 179.3 & 13.36\tablenotemark{d} \\
           & 2005.8455 & 10.24 &  84.3 & 10.46\tablenotemark{d} \\
           & 2005.8455 & 11.75 & 103.0 & 12.58\tablenotemark{d} \\
    GJ 281 & 2005.2286 &  5.74 & 237.0 & 12.48 \\
           & 2005.2286 &  8.80 & 288.4 & 12.92 \\
           & 2006.1158 & 10.64 & 224.6 & 13.43\tablenotemark{f} \\
    GJ 285 & 2005.2095 &  8.83 & 114.3 & 11.55 \\
  HD 75332 & 2005.3107 &  8.25 & 141.7 & 11.49 \\
  HD 82443 & 2004.9829 &  5.27 & 190.3 & 11.64\tablenotemark{d} \\
           & 2004.9829 &  5.42 & 191.5 & 16.65\tablenotemark{d} \\
           & 2004.9829 &  8.33 &  97.8 & 12.24\tablenotemark{d} \\
           & 2004.9829 & 10.17 & 164.8 & 16.14\tablenotemark{d} \\
           & 2004.9829 & 13.74 & 215.0 & 14.67\tablenotemark{d} \\
  HD 90905 & 2005.2098 &  5.47 & 188.2 & 10.91 \\
           & 2005.2098 & 12.41 & 176.8 & 13.32 \\
  HD 92945 & 2005.3983 &  9.77 & 236.2 & 12.82 \\
  HD 93528 & 2005.3271 &  4.82 & 332.3 & 14.27\tablenotemark{d} \\
    GJ 402 & 2006.1273 & 12.46 & 324.0 & 10.80\tablenotemark{f} \\
           & 2005.3164 & 13.88 & 337.7 & 10.10 \\
  HD 96064 & 2005.2972 &  4.69 &  29.7 & 15.99\tablenotemark{e} \\
           & 2005.2972 &  5.94 & 213.8 &  8.18 \\
           & 2005.2972 &  6.11 & 213.6 & 10.76 \\
           & 2005.2972 &  8.90 & 329.7 & 15.90 \\
 HD 102195 & 2005.3109 & 11.94 & 185.5 & 13.82 \\
 HD 102392 & 2005.3082 &  5.72 &  42.0 & 15.02 \\
           & 2005.3082 & 10.57 & 308.9 & 14.99\tablenotemark{e} \\
HD 108767B & 2005.3055 &  6.72 &  87.7 & 12.41 \\
           & 2005.3055 &  8.28 & 100.1 & 14.62 \\
           & 2005.3055 & 10.20 & 123.9 & 15.10 \\
 HD 109085 & 2005.3986 & 12.92 & 256.2 & 15.80 \\
BD+60 1417 & 2005.2946 &  2.05 & 298.4 &  8.76 \\
           & 2005.2946 & 14.08 & 133.5 & 12.80\tablenotemark{e} \\
 HD 116956 & 2005.4067 &  9.34 &  17.4 & 15.05 \\
  GJ 524.1 & 2005.2948 &  7.59 &  19.7 & 13.09 \\
 HD 124106 & 2005.2975 &  7.51 & 124.8 & 13.85 \\
           & 2005.2975 &  9.39 & 342.2 &  9.51 \\
           & 2005.2975 &  9.60 & 341.1 &  8.71 \\
           & 2005.2975 & 10.39 & 287.5 & 14.65 \\
           & 2005.2975 & 11.17 & 291.7 & 14.18 \\
           & 2005.2975 & 12.06 & 120.6 & 15.45 \\
 HD 130322 & 2005.4014 &  7.61 & 329.8 & 11.00 \\
 HD 130948 & 2005.2922 &  2.60 & 103.1 &  8.56\tablenotemark{g} \\
           & 2005.2922 &  2.66 & 104.0 &  8.83\tablenotemark{g} \\
 HD 135363 & 2005.2949 &  7.50 & 122.1 & 10.28 \\
 HD 139813 & 2005.4097 &  6.85 & 271.3 & 14.48\tablenotemark{d} \\
           & 2005.4097 &  7.36 & 272.2 & 14.79\tablenotemark{d} \\
 HD 141272 & 2005.2977 &  2.31 &  12.3 & 11.44 \\
           & 2005.2977 &  4.03 & 286.9 & 16.59\tablenotemark{e} \\
           & 2005.2977 &  8.04 & 305.4 & 15.78\tablenotemark{e} \\
           & 2005.2977 &  8.32 & 258.1 & 15.66 \\
           & 2005.2977 & 10.95 & 299.2 &  9.90 \\
           & 2005.2977 & 11.43 & 190.4 & 15.14 \\
           & 2005.2977 & 12.26 & 209.5 & 12.31 \\
    GJ 628 & 2005.2924 &  5.19 & 259.0 & 16.32\tablenotemark{e} \\
           & 2005.2924 &  6.29 & 161.6 & 14.81 \\
           & 2005.2924 & 10.25 & 232.1 & 14.90 \\
           & 2005.2924 & 10.52 &   2.8 & 13.42\tablenotemark{e} \\
           & 2005.2924 & 11.49 & 308.6 & 14.88 \\
           & 2005.2924 & 12.72 & 228.6 & 15.10 \\
           & 2005.2924 & 13.64 & 215.4 & 15.03 \\
 HIP 81084 & 2005.2979 &  6.84 & 234.6 & 13.10 \\
           & 2005.2979 &  8.64 &   4.5 & 13.70 \\
           & 2005.2979 &  9.69 &  49.6 & 11.22 \\
           & 2005.2979 &  9.98 &  68.4 & 13.04 \\
           & 2005.2979 & 14.04 & 226.4 & 11.60 \\
 HD 160934 & 2005.2952 &  4.08 & 319.2 & 11.03 \\
           & 2005.2952 &  8.94 & 232.9 &  9.96 \\
 HD 162283 & 2005.3007 &  2.76 & 118.7 & 15.54 \\
           & 2005.3007 &  3.41 & 154.4 & 14.13 \\
           & 2005.3007 &  3.26 &   4.9 & 15.21 \\
           & 2005.3007 &  3.68 & 244.8 & 14.01 \\
           & 2005.3007 &  3.94 & 152.8 & 14.22 \\
           & 2005.3007 &  4.22 & 158.8 & 12.52 \\
           & 2005.3007 &  4.47 & 111.9 & 15.21 \\
           & 2005.3007 &  4.37 &  23.1 & 13.23 \\
           & 2005.3007 &  4.70 & 299.8 & 15.44 \\
           & 2005.3007 &  4.95 &  80.0 & 10.82 \\
           & 2005.3007 &  5.24 & 124.0 & 16.08 \\
           & 2005.3007 &  5.24 & 191.8 & 13.38 \\
           & 2005.3007 &  5.23 & 348.9 & 15.59 \\
           & 2005.3007 &  6.07 &  72.2 & 14.49 \\
           & 2005.3007 &  6.61 & 159.4 & 12.06 \\
           & 2005.3007 &  6.72 &  84.6 & 14.28 \\
           & 2005.3007 &  6.89 & 352.8 & 13.50 \\
           & 2005.3007 &  7.11 &  91.2 &  8.65 \\
           & 2005.3007 &  7.38 & 176.6 & 16.29 \\
           & 2006.7069 &  7.42 &  18.0 & 15.95\tablenotemark{f} \\
           & 2005.3007 &  7.33 &  78.7 & 15.73 \\
           & 2005.3007 &  7.47 &  75.0 & 12.53 \\
           & 2005.3007 &  7.71 & 134.8 & 12.62 \\
           & 2005.3007 &  7.94 & 110.5 & 15.55 \\
           & 2005.3007 &  8.19 & 198.2 & 15.44 \\
           & 2005.3007 &  8.37 & 173.6 & 15.88 \\
           & 2005.3007 &  8.60 & 243.7 & 14.96 \\
           & 2005.3007 &  8.69 & 132.5 & 15.01 \\
           & 2006.7069 &  9.01 & 318.9 & 15.97\tablenotemark{f} \\
           & 2005.3007 &  9.08 & 141.2 & 12.99 \\
           & 2005.3007 &  9.23 & 152.7 & 15.98 \\
           & 2005.3007 &  9.32 & 142.4 & 14.95 \\
           & 2005.3007 &  9.37 & 205.7 & 13.88 \\
           & 2005.3007 &  9.28 & 337.0 & 12.77 \\
           & 2005.3007 &  9.63 & 235.2 & 15.38 \\
           & 2005.3007 &  9.56 &  78.8 & 13.78 \\
           & 2006.7069 &  9.86 & 286.6 & 15.74\tablenotemark{f} \\
           & 2006.7069 & 10.03 & 289.1 & 15.75\tablenotemark{f} \\
           & 2005.3007 & 10.20 &  97.7 & 13.31 \\
           & 2005.3007 & 10.18 &  37.3 &  8.81 \\
           & 2005.3007 & 10.31 & 319.4 & 14.47 \\
           & 2005.3007 & 10.48 & 308.7 & 15.38 \\
           & 2005.3007 & 10.71 & 337.2 & 14.52 \\
           & 2005.3007 & 10.92 &  36.3 & 11.98\tablenotemark{e} \\
           & 2005.3007 & 11.36 & 249.0 & 13.24 \\
           & 2005.3007 & 11.38 & 112.7 &  9.04 \\
           & 2005.3007 & 11.80 &  40.1 & 14.79\tablenotemark{e} \\
           & 2006.7069 & 12.00 & 261.7 & 15.23\tablenotemark{f} \\
           & 2005.3007 & 12.01 & 329.6 &  9.78 \\
           & 2005.3007 & 12.28 & 153.9 & 14.26 \\
           & 2006.7069 & 12.28 & 265.3 & 13.99\tablenotemark{f} \\
           & 2005.3007 & 12.43 & 114.8 &  9.89\tablenotemark{e} \\
           & 2006.7069 & 12.47 &  71.5 & 14.76\tablenotemark{f} \\
           & 2005.3007 & 12.40 & 100.7 & 13.16 \\
           & 2005.3007 & 12.61 & 111.8 & 13.37\tablenotemark{e} \\
           & 2005.3007 & 13.04 &  16.9 & 14.10\tablenotemark{e} \\
           & 2006.7069 & 13.21 & 161.5 & 13.02\tablenotemark{f} \\
           & 2006.7069 & 13.46 &  65.5 & 14.51\tablenotemark{f} \\
           & 2005.3007 & 13.53 & 303.6 & 12.68\tablenotemark{e} \\
           & 2006.7069 & 13.59 & 257.1 & 11.76\tablenotemark{f} \\
           & 2005.3007 & 14.12 & 314.4 & 12.55\tablenotemark{e} \\
           & 2005.3007 & 14.59 & 112.9 & 10.20\tablenotemark{e} \\
           & 2005.3007 & 14.66 &  17.4 & 12.43\tablenotemark{e} \\
           & 2005.3007 & 14.74 &  34.6 & 11.81\tablenotemark{e} \\
 HD 166181 & 2005.2925 & 10.38 &  53.4 & 14.40 \\
           & 2005.2925 & 11.21 & 195.8 & 15.04 \\
           & 2005.2925 & 13.40 & 167.6 & 14.19\tablenotemark{e} \\
           & 2005.2925 & 14.46 & 262.8 & 11.93\tablenotemark{e} \\
 HD 187748 & 2005.3965 &  5.51 & 325.9 & 15.76 \\
           & 2005.3965 &  7.93 & 277.1 & 13.01 \\
           & 2005.3965 &  8.02 & 276.7 & 12.18 \\
           & 2005.3965 & 12.81 & 114.3 &  9.74 \\
           & 2005.3965 & 13.15 & 321.5 & 12.52 \\
           & 2006.7043 & 15.02 & 311.3 & 14.90\tablenotemark{f} \\
  GJ 791.3 & 2005.3992 &  1.98 & 341.2 & 12.03\tablenotemark{d} \\
           & 2005.3992 &  2.39 &  51.3 & 13.51\tablenotemark{d} \\
           & 2005.3992 &  3.77 & 289.0 & 10.92\tablenotemark{d} \\
           & 2005.3992 &  3.80 & 137.6 & 13.99\tablenotemark{d} \\
           & 2005.3992 &  3.87 &  19.4 & 10.01\tablenotemark{d} \\
           & 2005.3992 &  4.38 & 201.6 & 14.31\tablenotemark{d} \\
           & 2005.3992 &  5.55 & 300.0 & 12.72\tablenotemark{d} \\
           & 2005.3992 &  5.96 &  49.6 & 10.28\tablenotemark{d} \\
           & 2005.3992 &  6.56 & 232.8 & 12.91\tablenotemark{d} \\
           & 2005.3992 &  6.66 & 155.9 & 13.37\tablenotemark{d} \\
           & 2005.3992 &  6.73 & 254.7 & 12.93\tablenotemark{d} \\
           & 2005.3992 &  8.01 &  10.3 & 13.03\tablenotemark{d} \\
           & 2005.3992 &  8.25 & 143.8 & 15.03\tablenotemark{d} \\
           & 2005.3992 &  8.31 &  71.5 & 15.02\tablenotemark{d} \\
           & 2005.3992 &  8.36 & 155.7 & 13.62\tablenotemark{d} \\
           & 2005.3992 &  8.89 & 177.3 & 12.86\tablenotemark{d} \\
           & 2005.3992 &  9.33 &  10.5 & 14.93\tablenotemark{d} \\
           & 2005.3992 &  9.55 & 276.2 & 15.45\tablenotemark{d} \\
           & 2005.3992 &  9.64 & 195.5 & 14.14\tablenotemark{d} \\
           & 2005.3992 &  9.89 & 347.0 & 15.16\tablenotemark{d} \\
           & 2005.3992 & 10.05 & 255.1 & 14.63\tablenotemark{d} \\
           & 2005.3992 & 10.12 & 201.2 & 13.28\tablenotemark{d} \\
           & 2005.3992 & 10.21 & 310.6 & 14.36\tablenotemark{d} \\
           & 2005.3992 & 10.22 & 328.8 & 10.56\tablenotemark{d} \\
           & 2005.3992 & 10.55 & 166.9 & 15.14\tablenotemark{d} \\
           & 2005.3992 & 10.63 &  80.3 & 13.51\tablenotemark{d} \\
           & 2005.3992 & 10.75 & 326.6 & 11.48\tablenotemark{d} \\
           & 2005.3992 & 10.80 &  57.8 & 10.86\tablenotemark{d} \\
           & 2005.3992 & 10.84 &  51.7 & 14.67\tablenotemark{d} \\
           & 2005.3992 & 11.26 & 243.5 &  9.22\tablenotemark{d} \\
           & 2005.3992 & 11.58 & 315.9 &  9.96\tablenotemark{d} \\
           & 2005.3992 & 11.71 &  14.8 & 13.26\tablenotemark{d} \\
           & 2005.3992 & 12.17 &  18.6 & 11.27\tablenotemark{d} \\
           & 2005.3992 & 12.45 &  46.6 & 12.52\tablenotemark{d} \\
           & 2005.3992 & 12.80 & 274.4 & 11.86\tablenotemark{d} \\
           & 2005.3992 & 13.07 & 127.8 & 11.44\tablenotemark{d} \\
           & 2005.3992 & 13.08 &  75.4 & 13.58\tablenotemark{d} \\
           & 2005.3992 & 13.23 &  81.0 & 11.26\tablenotemark{d} \\
           & 2005.3992 & 13.73 &  70.5 &  8.91\tablenotemark{d} \\
           & 2005.3992 & 14.41 &  65.8 & 11.04\tablenotemark{d} \\
           & 2005.3992 & 15.14 & 356.5 & 11.98\tablenotemark{d} \\
           & 2005.3992 & 15.23 & 179.8 & 12.02\tablenotemark{d} \\
 HD 201651 & 2005.4867 &  3.67 & 201.4 & 12.48 \\
           & 2005.4867 &  8.39 & 259.1 & 12.90 \\
           & 2005.4867 & 14.53 & 331.5 & 13.75\tablenotemark{e} \\
 HD 202575 & 2005.5386 &  5.54 &  28.5 & 12.41 \\
           & 2005.5386 & 12.37 & 168.0 & 13.98 \\
   GJ 4199 & 2004.6431 &  9.16 & 319.9 & 12.24 \\
           & 2004.6431 & 11.76 & 177.6 & 10.58 \\
 HD 206860 & 2005.6069 &  3.67 &  60.0 & 15.13 \\
 HD 208313 & 2005.4868 &  2.93 &  30.6 & 14.78 \\
           & 2005.4868 &  6.24 &  31.1 &  9.69 \\
           & 2005.4868 &  9.45 & 301.0 & 14.41 \\
           & 2005.4868 & 10.45 & 137.8 & 16.50\tablenotemark{e} \\
           & 2005.4868 & 11.43 & 151.6 & 15.85 \\
           & 2005.4868 & 13.23 & 121.0 & 13.73 \\
           & 2005.4868 & 13.51 &  33.2 & 14.61 \\
           & 2005.4868 & 15.13 &  63.8 & 12.00 \\
  V383 Lac & 2005.5660 &  4.00 & 100.0 & 14.14 \\
           & 2005.5660 &  4.02 &  79.3 & 15.98 \\
           & 2005.5660 &  4.63 & 204.6 & 11.81 \\
           & 2005.5660 &  4.70 & 207.7 & 14.89 \\
           & 2005.5660 &  8.49 & 181.6 & 14.82 \\
           & 2005.5660 &  9.09 & 110.0 & 11.65 \\
           & 2005.5660 &  9.55 & 358.6 & 14.15 \\
           & 2005.5660 & 10.59 &  93.0 &  8.48 \\
           & 2005.5660 & 11.68 & 142.3 & 11.24 \\
 HD 213845 & 2005.6453 & 12.85 & 214.2 & 14.91 \\
  GJ 875.1 & 2005.6071 &  7.83 & 343.9 &  9.40 \\
           & 2005.6071 &  8.97 & 260.7 & 12.73 \\
           & 2005.6071 & 11.42 & 151.3 & 12.30 \\
           & 2005.6071 & 14.54 &  15.2 & 11.85\tablenotemark{e} \\
   GJ 9809 & 2006.7019 &  2.10 & 240.4 & 15.30\tablenotemark{f} \\
           & 2005.5907 &  2.18 &  30.6 & 12.91 \\
           & 2006.7019 &  3.22 & 141.6 & 15.55\tablenotemark{f} \\
           & 2005.5907 &  3.35 & 207.8 & 12.31 \\
           & 2005.5907 &  3.55 & 337.2 & 11.53 \\
           & 2005.5907 &  3.79 & 173.8 & 11.92 \\
           & 2006.7019 &  4.51 &   0.9 & 16.23\tablenotemark{f} \\
           & 2005.5907 &  5.37 & 258.1 & 14.38 \\
           & 2006.7019 &  5.66 & 118.3 & 15.35\tablenotemark{f} \\
           & 2005.5907 &  6.39 &  68.9 & 15.47 \\
           & 2005.5907 &  7.01 & 101.1 &  9.69 \\
           & 2005.5907 &  7.34 & 236.9 & 13.49 \\
           & 2005.5907 &  7.69 & 127.3 & 11.26 \\
           & 2005.5907 &  7.75 & 131.1 & 14.21 \\
           & 2006.7019 &  7.92 &  78.4 & 15.65\tablenotemark{f} \\
           & 2006.7019 &  9.08 & 247.3 & 14.84\tablenotemark{f} \\
           & 2005.5907 &  9.18 &  36.5 & 15.03 \\
           & 2005.5907 &  9.23 &  27.2 & 14.80 \\
           & 2005.5907 &  9.51 &  95.2 & 12.35 \\
           & 2006.7019 &  9.58 &  84.1 & 15.30\tablenotemark{f} \\
           & 2005.5907 &  9.98 &  68.5 & 14.35 \\
           & 2005.5907 &  9.98 & 121.4 & 10.42 \\
           & 2006.7019 & 10.02 & 163.4 & 15.37\tablenotemark{f} \\
           & 2005.5907 & 10.18 &  93.9 & 13.69 \\
           & 2006.7019 & 10.73 &  12.2 & 15.33\tablenotemark{f} \\
           & 2005.5907 & 10.64 & 248.1 & 13.59 \\
           & 2005.5907 & 10.94 & 254.4 &  9.90 \\
           & 2005.5907 & 11.23 &  82.9 & 13.75 \\
           & 2006.7019 & 11.40 & 336.1 & 15.18\tablenotemark{f} \\
           & 2005.5907 & 11.69 & 155.9 & 12.14 \\
           & 2005.5907 & 11.87 &  32.2 & 12.28 \\
           & 2005.5907 & 11.74 & 291.2 &  7.01 \\
           & 2005.5907 & 12.23 & 310.2 &  8.73 \\
           & 2005.5907 & 12.75 &  66.9 & 12.76 \\
           & 2005.5907 & 13.03 &  58.0 & 11.09 \\
           & 2005.5907 & 12.93 & 332.5 & 13.42 \\
           & 2005.5907 & 14.04 & 309.7 & 12.50 \\
 HD 220140 & 2005.5934 &  6.14 & 358.5 & 15.96 \\
           & 2005.5934 & 15.19 &  50.4 & 10.62\tablenotemark{e} \\
 HD 221503 & 2005.6646 &  9.02 & 234.4 & 15.61\tablenotemark{d} \\
    GJ 900 & 2004.6458 &  7.41 &  76.0 & 14.20 \\
           & 2004.6458 & 12.15 & 150.6 & 12.53 \\
           & 2004.6458 & 12.41 &  96.4 &  9.36 \\
  GJ 907.1 & 2005.6837 &  7.93 & 296.7 & 13.68 \\
\enddata
\tablecomments{Target stars around which no point source was 
detected are omitted from this table. Unless stated otherwise,
all point sources listed were confirmed to be background objects
using data from two epochs.}
\tablenotetext{a}{Uncertainty is 0.015\arcsec, see text for detail.}
\tablenotetext{b}{Uncertainty is 0.2\degr, see text for detail.}
\tablenotetext{c}{Uncertainties are given in Table~\ref{tbl:sigphot}, see text for detail.}
\tablenotetext{d}{No second epoch data available.}
\tablenotetext{e}{Source undetected in second epoch data.}
\tablenotetext{f}{Source detected in second epoch data only.}
\tablenotetext{g}{Previously known brown dwarf companion \citep{potter02,goto02}.}
\end{deluxetable}
\twocolumngrid

\subsection{Multiple systems} \label{sect:binaries}

As mentioned in \S\ref{sect:sample}, 16 of the target stars are part of multiple systems. As an orbital solution can potentially be determined in a reasonable amount of time for close-separation multiple systems, the measured properties of the systems observed that have a separation below 2\arcsec\ are presented in Table~\ref{tbl:bin} as a reference for future studies. Two of the close-separation systems observed were resolved for the first time by our observations (HD~14802 and HD~166181), and a third system (HD~213845) was found to be a relatively large separation ($\sim$6\arcsec) binary for which we have found no prior indication in the literature; since HD~213845 is reported to be part of a binary system for the first time in this paper, its properties are presented in Table~\ref{tbl:bin} as well. The three multiple systems observed for the first time in this work are discussed in more detail below.

\paragraph{HD~14802} A source $12\pm2$ times fainter than HD~14802 was detected at a separation of $0.469\arcsec\pm0.005\arcsec$ and P.A. $267.1\degr\pm0.7\degr$ (epoch 2005.6348); the large uncertainty on the flux ratio is due to the peak of the primary star PSF being saturated. Common proper motion of the pair was not verified but the system is likely bound given the brightness and close separation of the companion. The Hipparcos catalog \citep{hipparcos} indicates that the proper motion of this star is accelerating and the star is likely part of a binary system; an astrometric solution for the system was obtained by \citet{gontcharov00}. The estimated period and semi-major axis are 25~yr and 0.5\arcsec, respectively, consistent with the projected separation we have measured. 

\begin{figure}
\epsscale{1}
\plotone{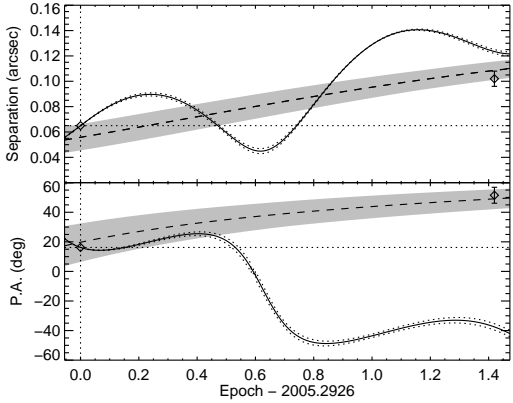}
\caption{\label{fig:pmhd166181} Verification of the physical association of the point source detected around HD~166181. Open diamonds mark the observed separation ({\it top}) and P.A. ({\it bottom}) of the point source at the two epochs. The solid line indicates the expected separation and P.A. of a distant background source as a function of time. The predicted separation and P.A. of HD~166181B based on the astrometric solution of \citet{fekel05} are shown as {\it dashed lines}, with uncertainties indicated by the shaded areas.}
\end{figure}

\paragraph{HD~166181} This star has been known for a long time to be a spectroscopic binary with a period of only 1.8 days \citep{nadal74}. More recently, analysis of additional radial velocity data has lead \citet{dempsey96} to propose that the system is in fact triple; a proposition which was confirmed by \citet{fekel05}, who found radial velocity variations ascribable to a third component with an orbit of period 5.7 year and eccentricity 0.765. Further, by reanalyzing Hipparcos data in light of this new component, these authors have found a new astrometric solution for the system, leading to revised values of parallax and proper motion (see Table~\ref{tbl:targets}) and to a determination of the orbital inclination of the long-period companion. Based on their complete solution, they estimate the semi-major axis of the outer companion at 0.077\arcsec\ (2.5~AU) and its mass at 0.79~$M_\odot$. Our observations have resolved the long-period companion of this triple system. In 2005.2926, the companion was located at a separation of $0.065\arcsec\pm 0.005\arcsec$ and P.A. of $16.2\degr\pm5.0\degr$, and in 2006.7124, it was located at a separation of $0.102\arcsec\pm 0.003\arcsec$ and P.A. of $51.5\degr\pm2.0\degr$. The evolution of the separation and P.A. of this source between the two epochs is far from that expected for an unrelated background source and is in very good agreement with the orbital motion expected based on the astrometric solution of \citet{fekel05} (see Fig.~\ref{fig:pmhd166181}), confirming that the source observed is HD~166181B. The flux ratio of the component Aab to component B is $\sim$5.5, a contrast of $\sim$1.85~mag.

\onecolumngrid
\begin{deluxetable}{lccccc}
\tablewidth{0pt}
\tabletypesize{\scriptsize}
\tablecolumns{6}
\tablecaption{Properties of close separation multiple systems \label{tbl:bin}}
\tablehead{
\colhead{Name} & \colhead{Epoch} & \colhead{Separation} & \colhead{P.A.} & \colhead{Brightness} & \colhead{1$^{\rm st}$ spatially resolved} \\
\colhead{} & \colhead{} & \colhead{(\arcsec)} & \colhead{(\degr)} & \colhead{ratio\tablenotemark{a}} & \colhead{observation}
}
\startdata
HD~14802AB  & 2005.6348 & $0.469\pm0.005$   & $267.1\pm0.7$ & $12\pm2$ & This work \\
GJ~234AB    & 2005.8455 & $1.392\pm0.002$ & $44.5\pm0.1$    & $4.7\pm0.1$ & W. Baade (1955), c.f. \citet{gatewood03}\\
HD~77407AB  & 2005.3163 & $1.702\pm0.004$ & $355.6\pm0.1$   & $7.4\pm0.3$ & \citet{mugrauer04}\\
HD~102392AB & 2005.3083 & $1.134\pm0.002$ & $172.5\pm0.1$   & $7.8\pm0.3$ & \citet{rossiter55} \\
            & 2006.1929 & $1.137\pm0.002$ & $171.1\pm0.1$   & & \\
HD~129333AB & 2005.3004 & $0.766\pm0.002$ & $173.0\pm0.2$   & $14.0\pm0.5$ & \citet{metchev04} \\
HD~135363AB & 2005.2950 & $0.302\pm0.002$ & $129.9\pm0.5$   & $4.0\pm0.1$ & This work\tablenotemark{b} \\
            & 2006.1277 & $0.316\pm0.002$ & $131.8\pm0.5$   & & \\
HD~160934AB & 2005.2953 & $0.213\pm0.002$ & $268.5\pm0.7$   & $2.2\pm0.1$ & \citet{hormuth07}\\
            & 2006.7097 & $0.218\pm0.002$ & $271.3\pm0.7$   & & \\
HD~166181AabB& 2005.2926& $0.065\pm0.005$ & $16.2\pm5.0$    & $5.5\pm0.4$ & This work \\
            & 2006.7124 & $0.102\pm0.003$ & $51.5\pm2.0$    & & \\
HD~167605AB & 2005.4017 & $1.182\pm0.002$ & $46.9\pm0.1$    & $8.2\pm0.2$ & \citet{arribas98} \\ 
HD~213845AB & 2005.6453 & $6.09\pm0.03$   & $129.8\pm0.4$   & $\sim$125\tablenotemark{c} & This work \\
            & 2006.5108 & $6.09\pm0.03$   & $129.6\pm0.4$   & & \\
GJ~900AB    & 2004.6459 & $0.611\pm0.002$ & $334.3\pm0.2$   & $5.7\pm0.2$ & \citet{martin03}\\
            & 2005.6864 & $0.673\pm0.002$ & $338.8\pm0.2$   & & \\
GJ~900AC    & 2004.6459 & $0.733\pm0.002$ & $344.7\pm0.2$   & $12.0\pm0.2$ & \citet{martin03}\\
            & 2005.6864 & $0.722\pm0.002$ & $345.6\pm0.2$   & & \\
GJ~907.1AB  & 2005.6837 & $0.789\pm0.002$ & $213.5\pm0.2$   & $1.62\pm0.05$ & \citet{rossiter55} \\ 
            & 2006.5411 & $0.775\pm0.002$ & $212.1\pm0.2$   & &
\enddata
\tablenotetext{a}{Brightness of the primary over that of the companion, in the NIRI CH4-short filter.}
\tablenotetext{b}{Independently resolved by \citet{biller07}, evidence for co-motion is reported in this work for the first time.}
\tablenotetext{c}{The peak of both the primary and companion is saturated in the data; the ratio quoted is an estimate based on the comparison of radial profiles.}
\end{deluxetable}
\twocolumngrid

\begin{figure}[t]
\epsscale{1}
\plotone{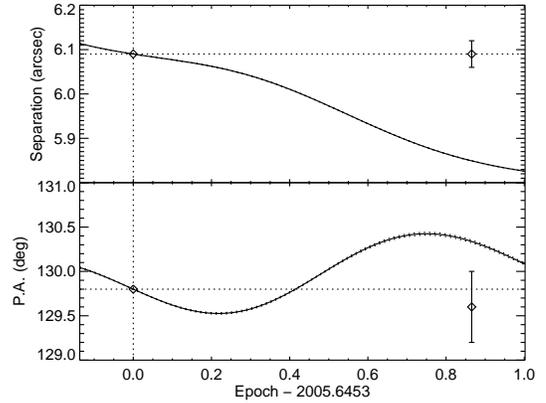}
\caption{\label{fig:pmhd213845} Same as Figure~\ref{fig:pmhd166181} for HD~213845}
\end{figure}

\paragraph{HD~213845} A bright source is visible in our data at a separation of $6.09\arcsec\pm0.03\arcsec$ and P.A. of $129.8\degr\pm0.4\degr$ from HD~213845 (epoch 2005.6453). This source did not change separation nor P.A. between our 2005 and 2006 observations (see Figure~\ref{fig:pmhd213845}), indicating that it is bound to HD~213845. The companion is only visible in our saturated data as its separation exceeds the field of view of the sub-array used for the unsaturated observations. Further, being relatively bright, the peak of the companion's PSF is saturated in all our data, making it very difficult to estimate its flux ratio to the primary and explaining the larger uncertainty on the separation and P.A. quoted above. We have nevertheless estimated that the companion is $\sim$125 times fainter than its primary based on a comparison of their radial intensity profiles at radii where the data are in the linear regime of the detector. The companion was possibly detected by 2MASS, but its measured position and photometry in the 2MASS point source catalog (PSC) are affected by confusion due to the nearby primary. Nevertheless, the relative position of this source in the 2MASS PSC, separation of 5.55\arcsec\ and P.A. of 128\degr, is consistent with the star being gravitationally bound to HD~213845 as, were it not a bound companion, its separation should have changed by $\sim$2\arcsec\ between the 2MASS observations and our first epoch observations. Although the separation of this binary system is well above the resolution limit of seeing-limited observations, we have found no prior indication of binarity in the literature.

\section{Analysis and discussion}\label{sect:discussion}

The detection limits determined in \S\ref{sect:limits} can be used to calculate an upper limit to the fraction of stars that have companions of mass and semi-major axis inside some given intervals. The analysis presented in this section is largely guided by the work of \citet{brandeker06, carson06, allen05}; and \citet{sivia96}. The statistical formalism for the analysis is presented first and various applications to our data are presented afterward.

\subsection{Statistical formalism}\label{sect:stat}

Consider the observation of $N$ stars enumerated by $j=1\ldots N$. Let $f$ be the fraction of stars that have at least one companion of mass and semi-major axis in the intervals $[\mmin, \mmax]$ and $[\amin, \amax]$, respectively, and $p_j$ the probability that such a companion around star $j$, if indeed it was there, would be detected given the detection limits of the observations. The probability of detecting such a companion around star $j$ is $fp_j$, and the probability of not detecting a companion around this star is simply $1-fp_j$. If the set $\{d_j\}$ denotes the detections made by the observations, such that $d_j$ equals 1 if a companion is detected around star $j$ or else equals 0, then the probability that the observed outcome would occur, also called the likelihood of the data given $f$, is given by
\begin{equation}\label{eq:lkhd}
\lkhd(\{d_j\} | f)=\prod_{j=1}^N \left( 1-fp_j \right)^{\left(1-d_j\right)} \left( fp_j \right)^{d_j}.
\end{equation}
According to Bayes' theorem, from the {\it a priori} probability density $p(f)$, or prior distribution, and the likelihood function \lkhd, one may calculate $p(f | \{d_j\})$, the probability density updated in light of the data, or posterior distribution:
\begin{equation}
p(f | \{d_j\}) = \frac{\lkhd(\{d_j\} | f) p(f)}{ \int_0^1 \lkhd(\{d_j\} | f) p(f) \drm f }.
\end{equation}
In this study, since we have no prior knowledge about $f$, we use the most ignorant prior distribution $p(f)=1$.

The posterior distribution $p(f | \{d_j\})$ can be used to determine a credible interval (CI) for $f$, bounded by \fmin\ and \fmax, for a given level of credibility $\alpha$. For a case where there is no detection, as is the case with our survey, then clearly $f_{\rm min}=0$, and the upper bound of the CI is found by solving
\begin{equation}\label{eq:cinodet}
\alpha=\int_0^{\fmax} p(f | \{d_j\}) \drm f.
\end{equation}
For a case where there are some detections, an equal-tail CI is found by solving
\begin{equation}\label{eq:cidet}
\frac{1-\alpha}{2}=\int_0^{\fmin} p(f | \{d_j\}) \drm f\quad {\rm and } \quad
\frac{1-\alpha}{2}=\int_{\fmax}^1 p(f | \{d_j\}) \drm f.
\end{equation}
In this work a value of $\alpha=0.95$ was chosen. 

The determination of the $p_j$'s is a critical step of this analysis; their value depends on the detection limits of the observations, on the ages and distances of the systems, and on the mass, semi-major axis, and orbital eccentricity distributions of the companions. In calculating the $p_j$'s it is also important to account properly for orbital inclination and phase as these affect significantly the distribution of projected separations for an orbit of given semi-major axis. In this work, the $p_j$'s were calculated using a Monte Carlo approach. The mass and semi-major axis intervals, $[\mmin, \mmax]$ and $[\amin, \amax]$, were first selected. Then for each target star, 10000 planets were generated by sampling randomly, for each planet, the mass, semi-major axis, orbital eccentricity, orbital separation projection factor, age of the system, and underlying residual noise in the image. The mass and semi-major axis distributions are left arbitrary for the moment; different possibilities will be explored later. For all of our calculations, the orbital eccentricity distribution was assumed to be that of the radial velocity exoplanets sample, which was approximated by a Gaussian function of mean 0.25, standard deviation 0.19, and with $0 \leq e \leq 0.8$ \citep{marcy05}. The orbital separation projection factor was sampled using the method described in Appendix~A of \citet{brandeker06}; this method properly takes into consideration orbital eccentricities, phases, and orientations. The age was sampled linearly within the range indicated in Table~\ref{tbl:targets}. Given the age assigned to each planet, the procedure described in \ref{sect:limits} was used to convert its mass into a magnitude difference in the NIRI CH4-short filter. The projected physical separation of each planet was converted into an angular separation based on the distance of its primary star. The random noise added to each planet was drawn from a Gaussian distribution of standard deviation equal to the measured residual noise at the separation of the planet. The effect of adding this noise is to either increase or decrease the signal that a planet would have in the residual image. The signal of some ``a priori detectable'' planets near the detection limit will thus decrease below the 5$\sigma$ detection limit, while the signal of some ``a priori undetectable'' planets will be boosted above the detection limit such that the appropriate detection completeness will result for planets of various true intensities (see \S\ref{sect:limits}). Finally, given the sample of planets assigned to target $j$, the probability $p_j$ was calculated as the fraction of planets lying above the corresponding 5$\sigma$ detection limits (c.f. Table~\ref{tbl:limits}).

The above determination of the $p_j$'s yields a CI for $f$ that is a function of the assumptions made on the mass and semi-major axis distributions. For a case where there is no detection, it is also possible to obtain a more conservative estimate of \fmax\ that is valid for any distributions of mass and semi-major axis. The procedure used to do this is identical to that described above except for the following. Rather than populating the whole intervals of mass and semi-major axis considered, all planets are assigned a mass and semi-major axis precisely equal to \mmin\ and \amin, respectively. Because more massive or more distant planets are easier to detect, the values of $p_j$'s calculated in this manner constitute lower limits to the values that would be obtained by populating the whole intervals assuming any specific distributions; accordingly, the resulting value of \fmax\ constitutes an upper limit. This approach is perfectly legitimate as long as \amax\ is chosen such that the values of $p_j$'s for any $a$ in $[\amin, \amax]$ are at least as large as those for \amin.

It is possible to derive a simple analytic expression for \fmax\ for a case where there is no detection; this expression may be useful to estimate what the results of an ongoing survey will be or scale actual results for different values of $N$ or detection probabilities. This expression may be obtained by first replacing each $p_j$ by the average detection probability $\left<p_j\right>$ in Eq.~(\ref{eq:lkhd}), and then recognizing that the likelihood function can be approximated by $e^{-Nf\left<p_j\right>}$. This leads to
\begin{equation}\label{eq:approx}
\fmax \approx \frac{-\ln\left(1-\alpha\right)}{N\left<p_j\right>}.
\end{equation}
This approximation, valid for $N\left<p_j\right> \gg 1$, is equivalent to using Poisson statistics rather than Binomial statistics for the presence of companions (c.f. Eq.~3--7 of \citealt{carson06}).

\subsection{\fmax\ for arbitrary mass and semi-major axis distributions}

\begin{figure}
\epsscale{1}
\plotone{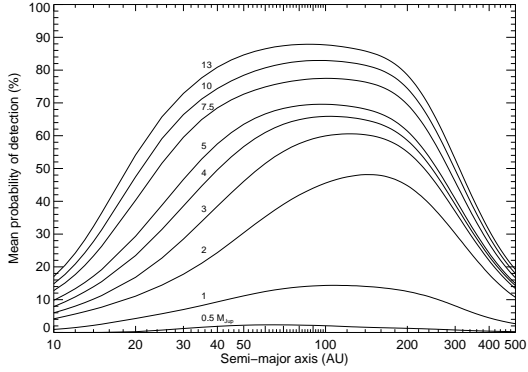}
\caption{\label{fig:prob} Mean probability of detection of a planet of given mass as a function of the semi-major axis of its orbit; the curves are labeled by the mass of the planet, in \mjup. The mean is obtained over all targets of the survey.}
\end{figure}

\begin{figure}
\epsscale{1}
\plotone{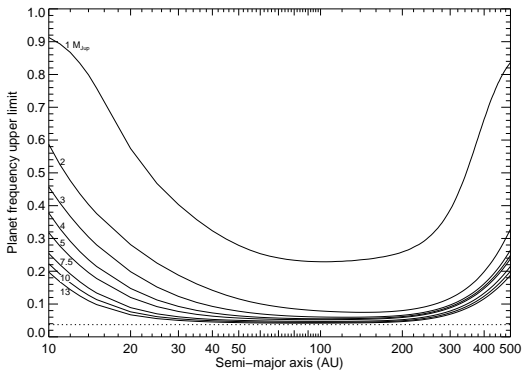}
\caption{\label{fig:fmax} Upper limits, with a credibility of 95\%, on the fraction of stars harboring at least one companion of mass in the range $[\mmin, 40]$~\mjup\ and orbit of semi-major axis in various ranges. The minimum mass, \mmin, is indicated on each curve. For any interval, $[\amin, \amax]$~AU, of semi-major axis selected, the correct value of \fmax\ to read from the graph is the maximum of the curve within that interval. The curves shown in this graph are conservative upper limits that are valid for any distributions of mass and semi-major axis. The dotted line indicates the minimum upper limit that one could derive from observation of 79 stars if the probability of detection of a planet was 100\% irrespective of its age, mass, and orbital separation.}
\end{figure}

As a first analysis of the survey results, we present estimates of \fmax\ that are independent of the mass and semi-major axis distributions for $\mmin=$0.5, 1, 2, 3, 4, 5, 7.5, 10, and 13~\mjup, and for all \amin\ between 10 and 500~AU; these estimates were calculated according to the last procedure described above. For this analysis, and those in the next section, we have not considered the 6 stars with candidates for which second epoch observations are missing. The results obtained in this section are valid for any \mmax\ up to $\sim$40~\mjup\ as no companion with a mass below this value was detected.\footnote{The previously known 40--65~\mjup\ binary brown dwarf companion located 2.6\arcsec from HD~130948 \citep{potter02, goto02} is detected in our data.} Planet detection probabilities for each star are indicated in Table~\ref{tbl:probs} for a small selection of masses and semi-major axes, while the mean planet detection probabilities, i.e. the average of the $p_j$'s over all $j$'s, are shown in Figure~\ref{fig:prob} as a continuous function of semi-major axis and for a larger selection of masses. The peak sensitivity of the survey occurs for semi-major axes between 50 and 200~AU; the peak values are $\sim$45\% and $\sim$68\% for 2 and 5~\mjup, respectively. The survey is particularly sensitive to brown dwarfs ($m\gtrsim 13$~\mjup), with a detection probability above 75\% between 35 and 200~AU. A decline in sensitivity occurs at a separation of $\sim$200~AU; this is consistent with the field of view of the observations ($\sim$11\arcsec\ radius) and mean distance of the targets (22~pc).

The results for \fmax\ are shown in Figure~\ref{fig:fmax}. For a semi-major axis interval lower bound of 50~AU, the 95\% credible planet frequency upper limits are 0.28 for 1--13~\mjup\ and 50--225~AU, 0.12 for 2--13~\mjup\ and 50--295~AU, and 0.057 for 5--13~\mjup\ and 50--185~AU. For a semi-major axis lower bound of 25~AU, the upper limits are 0.23 for 2--13~\mjup\ and 25--420~AU and 0.09 for 5--13~\mjup\ and 25--305~AU. For completeness, the exercise was repeated for circular orbits and for a uniform distribution of eccentricity (between 0 and 1), and the results obtained were very similar to those shown in Figure~\ref{fig:fmax}.

\begin{figure}
\epsscale{1}
\plotone{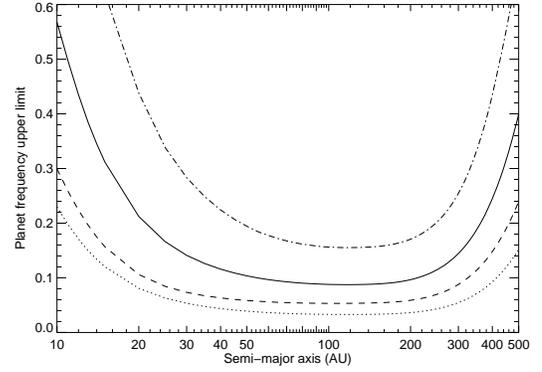}
\caption{\label{fig:fmax_m} Upper limits, with a credibility of 95\%, on the fraction of stars harboring at least one planet of mass in the range $[0.5,13]$~\mjup, assuming $\drm n/\drm m \propto m^\beta$, and semi-major axis in various ranges. The values of $\beta$ are $-2$ ({\it dot-dashed line}), $-1.2$ ({\it solid line}), and 0 ({\it dashed line}). For any interval, $[\amin, \amax]$~AU, of semi-major axis selected, the correct value of \fmax\ to read from the graph is the maximum of the curve within that interval. The 67\% credibility curve for $\beta=-1.2$ is also shown ({\it dotted line}).}
\end{figure}

\begin{figure*}
\epsscale{1}
\plotone{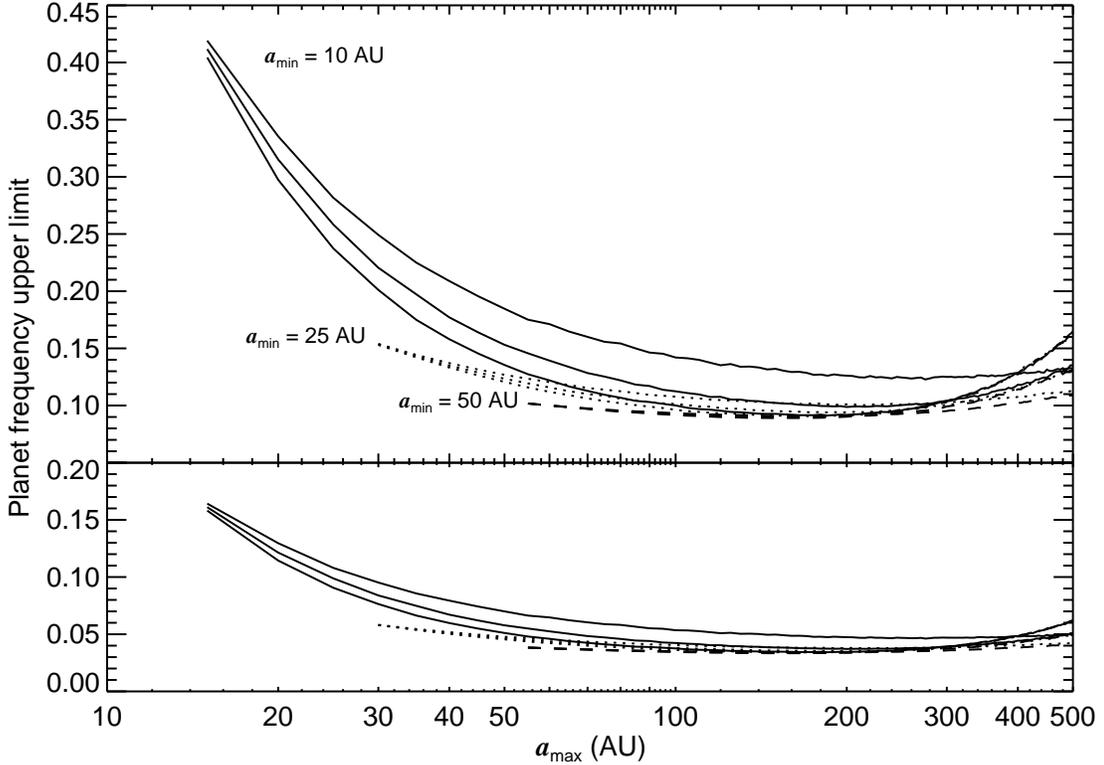}
\caption{\label{fig:fmax_ma} Upper limits, with a credibility of 95\% ({\it top panel}) or 67\% ({\it bottom panel}), on the fraction of stars harboring at least one giant planet of mass in the range $[0.5,13]$~\mjup, assuming $\drm n/\drm m \propto m^{-1.2}$, and orbit of semi-major axis in the range $[\amin, \amax]$~AU, assuming $\drm n/\drm a \propto a^\gamma$. The abscissa indicates the upper bounds (\amax) of the semi-major axis intervals, while the lower bounds (\amin) are 10~AU ({\it solid lines}), 25~AU ({\it dotted lines}), and 50~AU ({\it dashed lines}). The top, middle, and bottom curves in each set of three curves are for $\gamma=-1$, 0, and 1, respectively.}
\end{figure*}

The results also indicate that no more than 0.056 of stars have low-mass brown dwarf companions ($13 < m/M_{\rm Jup} <40$) between 25 and 250~AU. For determining the frequency of stars with at least one companion in the whole brown dwarf mass range ($13 < m/M_{\rm Jup} <75$) over the same range of semi-major axis, the brown dwarf companion to HD~130948 \citep{potter02, goto02} must be taken into account explicitly. This analysis must be carried out with care as the semi-major axis of this companion could be significantly different from its measured projected physical separation of 47~AU. It is possible to account for this uncertainty by calculating the probability distribution of the real semi-major axis of the brown dwarf companion using a Monte Carlo approach similar to the one presented above for the calculation of the $p_j$'s. Basically, the projected separation of the companion is fixed at $s=47$~AU and its orbital eccentricity and orbital projection factor are sampled randomly $10^5$ times, as described above. The de-projected semi-major axis is then calculated for each random trial and its normalized distribution over all trials is obtained. As the projection factor can never be larger than ($1+e_{\rm max}$), where $e_{\rm max}$ is the maximum eccentricity allowed, the semi-major axis probability distribution is equal to zero below $s/(1+e_{\rm max})$; the distribution extends to infinity for higher values. Applied to the current case, this calculation indicates that at a 95\% credible interval for the semi-major axis of the binary brown dwarf companion to HD~130948 is 26--157~AU. We thus posit that our observations have resulted in one detection in the semi-major axis interval 25--200~AU and mass interval 13--75~\mjup; then using the procedure described in the previous section and Eq.~(\ref{eq:cidet}), the 95\% credible interval for the frequency of stars with at least one brown dwarf companion in the semi-major axis interval 25--250~AU is $0.019_{-0.015}^{+0.083}$. This result is consistent with the upper limit of 0.12 (95\% credibility) reported by \citet{carson06} for the 25--100~AU semi-major axis interval and also with the fraction of $0.068_{-0.049}^{+0.083}$ (95\% credibility) reported by \citet{metchev04} for the range 30--1600~AU. For smaller semi-major axes, our results indicate that, with a credibility of 95\%, the fraction of stars with at least one brown dwarf companion in the range 10--25~AU is less than 0.20, and less than 0.10 for the range 15--25~AU.

\subsection{\fmax\ for specific mass and semi-major axis distributions}

In this section we derive first an upper limit to the fraction of stars harboring at least one planet in the single mass interval $[0.5,13]$~\mjup, assuming that the mass distribution follows $\drm n/\drm m \propto m^{-1.2}$. The mass distribution adopted is based on a statistical analysis of the RV results that properly accounts for the detection sensitivity reached for each star (A. Cumming et al. 2007, in preparation) and is formally valid only for planets with semi-major axis below $\sim$3~AU; here it is blindly extrapolated to larger semi-major axes. For comparison, a simple fit of the mass distribution of the RV exoplanets sample yields $\drm n/\drm m \propto m^{-1.1}$ \citep{butler06}. For this calculation the whole mass interval is populated according to the distribution stated, but all planets are assigned a value \amin\ for the semi-major axis, so as to make the results independent of its distribution. The calculation was made for all \amin\ between 10 and 500~AU. The results are shown in Figure~\ref{fig:fmax_m}. With a credibility of 95\%, the fraction of stars having at least one planet of mass in the range $[0.5,13]$~\mjup\ and semi-major axis in $[10,500]$, $[25,340]$, and $[50,230]$~AU is less than 0.57, 0.17, and 0.10, respectively. For reference, results of the same analysis assuming $\drm n/\drm m \propto m^{\beta}$, with $\beta=0$ and $-2$, are presented also in Figure~\ref{fig:fmax_m}. As expected, a smaller $\beta$ leads to larger values of \fmax\ because a larger fraction of planets have a smaller mass, while a larger value of $\beta$ has the opposite effect.

\begin{figure}
\epsscale{1}
\plotone{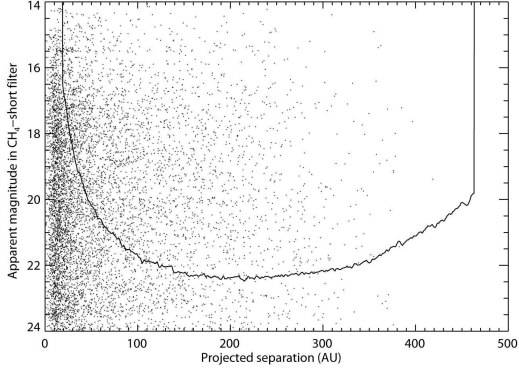}
\caption{\label{fig:mc_pl} Detection limits (5$\sigma$, {\it solid line}) and synthetic population of planets ({\it dots}) for the star HD~166181. A planet mass distribution following $dn/dm \propto m^{-1.2}$ inside 0.5--13~\mjup\ and a semi-major axis distribution following $dn/da \propto a^{-1}$ inside 10--300~AU were used. For this particular example, the planet detection probability $p_j$ is 30\%. }
\end{figure}

Next we calculate upper limits for the same mass interval by assuming further that the distribution of semi-major axes follows $\drm n/\drm a \propto a^{\gamma}$, for $\gamma=-1,$ 0, and 1. This range of power-law index includes the value of $\gamma=-0.67$ found by A. Cumming et al. (2007, in preparation) for the RV exoplanets sample within the range 0.03--3~AU. We have done the calculations for \amin=10, 25, and 50~AU, and for all \amax\ in the range $[\amin+5, 500]$~AU; the results are presented in Figure~\ref{fig:fmax_ma} and the corresponding planet detection probabilities for each star are shown in Table~\ref{tbl:probs} for a selection of semi-major axis intervals. Figure~\ref{fig:mc_pl} illustrates how a synthetic population of planets based on the above distributions compares with our detection limits. For $\gamma=-1$, the 95\% credible upper limits to the fraction of stars with at least one planet of mass in the range $[0.5, 13]$~\mjup\  are 0.28 for the semi-major axis range 10--25~AU, 0.18 for 10--50~AU, 0.13 for 25--50~AU, 0.11 for 25--100~AU, and 0.093 for 50--250~AU. Slightly smaller values of \fmax\ are found for larger values of $\gamma$, as such indices would place more planets at larger separations where they would have been easier to detect with our observations. For the larger values of \amin, the value of $\gamma$ has very little effect on the upper limit found as, irrespective of the value of $\gamma$, the majority of planets are located at separations for which the sensitivity of the observations is high. Overall, the weak dependence of \fmax\ on $\gamma$ implies that the semi-major axis distribution (i.e. $\gamma$) cannot be constrained from our results.

As one may worry that the population of planets around M dwarfs is different from that around earlier-type stars, because of smaller disk masses for example, we derive an estimate of \fmax\ by excluding the M dwarfs from the statistical analysis. This estimate is obtained using Eq.~(\ref{eq:approx}) and the values of the last three columns of Table~\ref{tbl:probs}; it is thus valid for $\beta=-1.2$ and $\gamma=-1$. Excluding M dwarfs from the sample leaves 64 stars and results in average detection probabilities $\left<p_j\right>$ of 0.070, 0.229, and 0.385 for 10--25~AU, 25--50~AU, and 50--250~AU, respectively. The corresponding 95\% credible upper limits to the fraction of stars with planets are then 0.67, 0.20, and 0.12. The effect is quite significant at the smallest orbital separations, where M dwarfs provide good sensitivities due to their smaller luminosity, smaller average distance, and younger average age. Similarly, the population of planets in stellar multiple systems may be different from that in single systems. Excluding multiples from the sample also leaves 64 stars and yields values of $\left<p_j\right>$ of 0.139, 0.299, and 0.389 for 10--25~AU, 25--50~AU, and 50--250~AU. The corresponding upper limits to the fraction of stars with planets are 0.34, 0.16, and 0.12; the effect is thus rather small in this case.

\onecolumngrid
\begin{deluxetable}{lccccccccccc}
\tablewidth{400pt}
\tabletypesize{\scriptsize}
\tablecolumns{12}
\tablecaption{Planet detection probability \label{tbl:probs}}
\tablehead{
 & \multicolumn{3}{c}{2~\mjup} & & \multicolumn{3}{c}{5~\mjup} & & \multicolumn{3}{c}{0.5--13~\mjup, $\beta=-1.2$, $\gamma=-1$} \\
\cline{2-4}
\cline{6-8}
\cline{10-12}
\colhead{Name} & \colhead{25~AU} & \colhead{50~AU} & \colhead{100~AU} & \colhead{ } & \colhead{25~AU} & \colhead{50~AU} & \colhead{100~AU} & \colhead{ } & \colhead{10--25~AU} & \colhead{25--50~AU} & \colhead{50--250~AU}
}
\startdata
      HD 166 & 0.002 & 0.261 & 0.748 &  & 0.693 & 0.936 & 0.985 &  & 0.120 & 0.339 & 0.448 \\
      HD 691 & 0.000 & 0.078 & 0.403 &  & 0.113 & 0.730 & 0.944 &  & 0.022 & 0.191 & 0.452 \\
     HD 1405 & 0.009 & 0.602 & 0.919 &  & 0.540 & 0.906 & 0.979 &  & 0.076 & 0.348 & 0.654 \\
     HD 5996 & 0.000 & 0.035 & 0.164 &  & 0.134 & 0.575 & 0.900 &  & 0.030 & 0.176 & 0.344 \\
     HD 9540 & 0.001 & 0.041 & 0.117 &  & 0.112 & 0.370 & 0.641 &  & 0.028 & 0.148 & 0.269 \\
    HD 10008 & 0.000 & 0.010 & 0.282 &  & 0.161 & 0.792 & 0.957 &  & 0.044 & 0.212 & 0.419 \\
       GJ 82 & 0.921 & 0.983 & 0.963 &  & 0.978 & 0.993 & 0.967 &  & 0.575 & 0.788 & 0.648 \\
    HD 14802 & 0.000 & 0.000 & 0.000 &  & 0.000 & 0.000 & 0.000 &  & 0.000 & 0.000 & 0.000 \\
    HD 16765 & 0.000 & 0.019 & 0.231 &  & 0.149 & 0.730 & 0.944 &  & 0.029 & 0.192 & 0.389 \\
    HD 17190 & 0.001 & 0.016 & 0.041 &  & 0.037 & 0.115 & 0.199 &  & 0.008 & 0.059 & 0.151 \\
    HD 17382 & 0.014 & 0.138 & 0.322 &  & 0.296 & 0.752 & 0.949 &  & 0.064 & 0.236 & 0.413 \\
    HD 17925 & 0.867 & 0.972 & 0.843 &  & 0.947 & 0.987 & 0.848 &  & 0.454 & 0.697 & 0.508 \\
    HD 18803 & 0.000 & 0.000 & 0.000 &  & 0.000 & 0.000 & 0.008 &  & 0.001 & 0.019 & 0.087 \\
    HD 19994 & 0.000 & 0.000 & 0.000 &  & 0.000 & 0.000 & 0.000 &  & 0.000 & 0.005 & 0.062 \\
    HD 20367 & 0.000 & 0.003 & 0.222 &  & 0.012 & 0.486 & 0.894 &  & 0.009 & 0.123 & 0.393 \\
      2E 759 & 0.001 & 0.197 & 0.648 &  & 0.396 & 0.872 & 0.973 &  & 0.089 & 0.282 & 0.490 \\
    HD 22049 & 0.000 & 0.000 & 0.000 &  & 0.935 & 0.338 & 0.068 &  & 0.273 & 0.257 & 0.035 \\
   HIP 17695 & 0.772 & 0.951 & 0.988 &  & 0.944 & 0.987 & 0.996 &  & 0.386 & 0.614 & 0.633 \\
    HD 25457 & 0.000 & 0.030 & 0.635 &  & 0.148 & 0.797 & 0.956 &  & 0.022 & 0.203 & 0.449 \\
   HD 283750 & 0.369 & 0.864 & 0.971 &  & 0.840 & 0.966 & 0.991 &  & 0.202 & 0.506 & 0.656 \\
      GJ 182 & 0.530 & 0.906 & 0.979 &  & 0.828 & 0.964 & 0.991 &  & 0.222 & 0.610 & 0.873 \\
      GJ 281 & 0.021 & 0.310 & 0.548 &  & 0.789 & 0.956 & 0.989 &  & 0.169 & 0.375 & 0.416 \\
      GJ 285 & 0.982 & 0.992 & 0.405 &  & 0.987 & 0.993 & 0.407 &  & 0.811 & 0.961 & 0.425 \\
    HD 72905 & 0.000 & 0.000 & 0.052 &  & 0.157 & 0.804 & 0.955 &  & 0.046 & 0.209 & 0.322 \\
    HD 75332 & 0.000 & 0.026 & 0.472 &  & 0.029 & 0.624 & 0.924 &  & 0.011 & 0.148 & 0.456 \\
    HD 77407 & 0.007 & 0.381 & 0.854 &  & 0.282 & 0.828 & 0.965 &  & 0.027 & 0.266 & 0.649 \\
    HD 78141 & 0.116 & 0.719 & 0.942 &  & 0.708 & 0.938 & 0.986 &  & 0.121 & 0.413 & 0.601 \\
    HD 82558 & 0.373 & 0.867 & 0.971 &  & 0.825 & 0.962 & 0.990 &  & 0.184 & 0.505 & 0.661 \\
      GJ 393 & 0.971 & 0.992 & 0.511 &  & 0.981 & 0.994 & 0.513 &  & 0.664 & 0.789 & 0.402 \\
    HD 90905 & 0.000 & 0.060 & 0.553 &  & 0.073 & 0.716 & 0.940 &  & 0.014 & 0.173 & 0.470 \\
    HD 91901 & 0.000 & 0.000 & 0.005 &  & 0.001 & 0.020 & 0.058 &  & 0.001 & 0.009 & 0.045 \\
    HD 92945 & 0.003 & 0.444 & 0.886 &  & 0.517 & 0.902 & 0.978 &  & 0.079 & 0.322 & 0.530 \\
    HD 93528 & 0.000 & 0.020 & 0.371 &  & 0.014 & 0.570 & 0.915 &  & 0.009 & 0.132 & 0.431 \\
      GJ 402 & 0.958 & 0.942 & 0.271 &  & 0.991 & 0.951 & 0.274 &  & 0.581 & 0.672 & 0.225 \\
    HD 96064 & 0.064 & 0.642 & 0.927 &  & 0.664 & 0.929 & 0.984 &  & 0.100 & 0.383 & 0.598 \\
    HD 97334 & 0.007 & 0.252 & 0.610 &  & 0.513 & 0.902 & 0.978 &  & 0.063 & 0.300 & 0.489 \\
   HD 102195 & 0.000 & 0.002 & 0.015 &  & 0.010 & 0.057 & 0.116 &  & 0.003 & 0.027 & 0.097 \\
   HD 102392 & 0.000 & 0.004 & 0.018 &  & 0.014 & 0.064 & 0.124 &  & 0.005 & 0.030 & 0.100 \\
   HD 105631 & 0.000 & 0.000 & 0.000 &  & 0.000 & 0.000 & 0.000 &  & 0.000 & 0.017 & 0.107 \\
   HD 107146 & 0.000 & 0.197 & 0.798 &  & 0.286 & 0.845 & 0.966 &  & 0.035 & 0.250 & 0.533 \\
  HD 108767B & 0.001 & 0.122 & 0.456 &  & 0.169 & 0.774 & 0.953 &  & 0.038 & 0.214 & 0.468 \\
   HD 109085 & 0.000 & 0.000 & 0.000 &  & 0.000 & 0.000 & 0.093 &  & 0.001 & 0.022 & 0.128 \\
  BD+60 1417 & 0.749 & 0.946 & 0.987 &  & 0.938 & 0.985 & 0.996 &  & 0.384 & 0.612 & 0.645 \\
   HD 111395 & 0.000 & 0.000 & 0.000 &  & 0.002 & 0.175 & 0.532 &  & 0.018 & 0.118 & 0.207 \\
   HD 113449 & 0.005 & 0.488 & 0.894 &  & 0.604 & 0.920 & 0.982 &  & 0.096 & 0.340 & 0.548 \\
    GJ 507.1 & 0.000 & 0.010 & 0.023 &  & 0.033 & 0.093 & 0.141 &  & 0.011 & 0.053 & 0.100 \\
   HD 116956 & 0.000 & 0.051 & 0.237 &  & 0.174 & 0.715 & 0.942 &  & 0.043 & 0.200 & 0.375 \\
   HD 118100 & 0.155 & 0.802 & 0.956 &  & 0.810 & 0.959 & 0.989 &  & 0.191 & 0.452 & 0.584 \\
    GJ 524.1 & 0.001 & 0.012 & 0.025 &  & 0.034 & 0.100 & 0.147 &  & 0.011 & 0.058 & 0.116 \\
   HD 124106 & 0.000 & 0.000 & 0.000 &  & 0.000 & 0.002 & 0.105 &  & 0.003 & 0.056 & 0.163 \\
  HD 125161B & 0.000 & 0.008 & 0.022 &  & 0.016 & 0.064 & 0.116 &  & 0.004 & 0.031 & 0.092 \\
   HD 129333 & 0.014 & 0.296 & 0.799 &  & 0.246 & 0.817 & 0.962 &  & 0.028 & 0.254 & 0.582 \\
   HD 130004 & 0.001 & 0.011 & 0.026 &  & 0.037 & 0.099 & 0.154 &  & 0.012 & 0.056 & 0.114 \\
   HD 130322 & 0.000 & 0.000 & 0.000 &  & 0.000 & 0.000 & 0.005 &  & 0.001 & 0.016 & 0.104 \\
   HD 130948 & 0.023 & 0.451 & 0.876 &  & 0.584 & 0.916 & 0.981 &  & 0.071 & 0.336 & 0.567 \\
   HD 135363 & 0.005 & 0.370 & 0.858 &  & 0.333 & 0.856 & 0.969 &  & 0.045 & 0.284 & 0.568 \\
   HD 141272 & 0.001 & 0.167 & 0.512 &  & 0.573 & 0.915 & 0.980 &  & 0.086 & 0.304 & 0.462 \\
  HD 147379B & 0.168 & 0.264 & 0.252 &  & 0.815 & 0.961 & 0.872 &  & 0.241 & 0.372 & 0.279 \\
      GJ 628 & 0.077 & 0.056 & 0.013 &  & 0.468 & 0.333 & 0.067 &  & 0.209 & 0.240 & 0.048 \\
   HIP 81084 & 0.002 & 0.364 & 0.863 &  & 0.494 & 0.898 & 0.977 &  & 0.085 & 0.315 & 0.528 \\
   HD 160934 & 0.500 & 0.900 & 0.978 &  & 0.855 & 0.968 & 0.992 &  & 0.231 & 0.574 & 0.780 \\
   HD 162283 & 0.002 & 0.019 & 0.035 &  & 0.049 & 0.125 & 0.200 &  & 0.016 & 0.072 & 0.140 \\
   HD 166181 & 0.000 & 0.058 & 0.508 &  & 0.119 & 0.756 & 0.949 &  & 0.021 & 0.193 & 0.468 \\
   HD 167605 & 0.000 & 0.003 & 0.015 &  & 0.004 & 0.043 & 0.093 &  & 0.002 & 0.018 & 0.078 \\
   HD 187748 & 0.000 & 0.126 & 0.689 &  & 0.181 & 0.799 & 0.958 &  & 0.036 & 0.223 & 0.516 \\
   HD 197481 & 0.910 & 0.980 & 0.808 &  & 0.913 & 0.980 & 0.810 &  & 0.538 & 0.897 & 0.622 \\
   HD 201651 & 0.000 & 0.004 & 0.015 &  & 0.008 & 0.043 & 0.089 &  & 0.002 & 0.019 & 0.073 \\
   HD 202575 & 0.009 & 0.102 & 0.204 &  & 0.262 & 0.651 & 0.921 &  & 0.070 & 0.218 & 0.333 \\
     GJ 4199 & 0.518 & 0.904 & 0.978 &  & 0.841 & 0.965 & 0.991 &  & 0.213 & 0.567 & 0.799 \\
   HD 206860 & 0.000 & 0.018 & 0.255 &  & 0.427 & 0.882 & 0.974 &  & 0.082 & 0.262 & 0.413 \\
   HD 208313 & 0.012 & 0.106 & 0.200 &  & 0.296 & 0.676 & 0.923 &  & 0.082 & 0.228 & 0.351 \\
    V383 LAC & 0.024 & 0.443 & 0.869 &  & 0.584 & 0.915 & 0.980 &  & 0.105 & 0.347 & 0.567 \\
   HD 213845 & 0.000 & 0.001 & 0.172 &  & 0.056 & 0.710 & 0.938 &  & 0.016 & 0.168 & 0.394 \\
    GJ 875.1 & 0.214 & 0.793 & 0.954 &  & 0.917 & 0.982 & 0.994 &  & 0.284 & 0.463 & 0.479 \\
      GJ 876 & 0.059 & 0.054 & 0.013 &  & 0.372 & 0.333 & 0.068 &  & 0.160 & 0.219 & 0.053 \\
     GJ 9809 & 0.811 & 0.960 & 0.990 &  & 0.920 & 0.982 & 0.994 &  & 0.396 & 0.717 & 0.893 \\
   HD 220140 & 0.150 & 0.736 & 0.945 &  & 0.771 & 0.952 & 0.988 &  & 0.136 & 0.426 & 0.573 \\
      GJ 900 & 0.008 & 0.443 & 0.874 &  & 0.644 & 0.928 & 0.984 &  & 0.116 & 0.353 & 0.491 \\
    GJ 907.1 & 0.000 & 0.000 & 0.007 &  & 0.009 & 0.528 & 0.906 &  & 0.015 & 0.134 & 0.344 \\
\hline
        Mean & 0.144 & 0.304 & 0.456 &  & 0.388 & 0.635 & 0.696 &  & 0.122 & 0.284 & 0.392
\enddata
\end{deluxetable}
\twocolumngrid

\section{Summary and conclusion}\label{sect:conclusion}

In this paper, we have presented the results of the Gemini Deep Planet Survey, a near-infrared adaptive optics search for giant planets on orbits of 10--300~AU around nearby young stars. The use of angular differential imaging at the Gemini North telescope has enabled us to reach the best sensitivities to date for detecting giant exoplanets with projected separations above $\sim$0.7\arcsec. The typical detection limits (5$\sigma$) reached by the survey, in magnitude difference between an off-axis point source and the central star, are 9.5 at 0.5\arcsec, 12.9 at 1\arcsec, 15 at 2\arcsec, and 16.5 at 5\arcsec, sufficient to detect planets more massive than 2~\mjup\ with a projected separation of 40--200~AU around a typical target star. More than 300 faint point sources have been detected around 54 of the 85 stars observed, but observations at a second epoch have revealed changes in separation and P.A. of these point sources relative to the target stars that are all consistent with those expected from unrelated background objects. The observations made as part of this survey have resolved the stars HD~14802, HD~166181, and HD~213845 into binaries for the first time.

We have presented a statistical analysis of the survey results to derive upper limits to the fraction of stars having planetary companions. This analysis indicates that the 95\% credible upper limit to the fraction of stars harboring at least one planet more massive than 2~\mjup\ with an orbit of semi-major axis in the range 25--420~AU or 50--295~AU is 0.23 or 0.12, respectively, independently of the mass and semi-major axis distributions of the planets; for planets more massive than 5~\mjup, the upper limits are 0.09 for 25-305~AU and 0.057 for 50--185~AU. It was also found that less than 0.056 of stars have low-mass brown dwarf companions ($13 < m/\mjup < 40$) between 25 and 250~AU (see Figure~\ref{fig:fmax}); this upper limit is set by the sample size only as the sensitivity of the observations to brown dwarfs is very good. Considering the whole brown dwarf mass range, the 95\% credible interval for the frequency of stars with at least one brown dwarf companion in the semi-major axis interval 25--250~AU is $0.019_{-0.015}^{+0.083}$. Assuming a mass distribution following $\drm n/\drm m \propto m^{-1.2}$, the results indicate that with a credibility of 95\% the fraction of stars having at least one planet of mass in the range 0.5--13~\mjup\ and semi-major axis in the range 25--325~AU is less than 0.17, and less than 0.10 for the range 50--220~AU. Assuming further a semi-major axis distribution following $\drm n/\drm a \propto a^{-1}$, the upper limits to the fraction of stars with planets are 0.28 for the range 10--25~AU, 0.13 for 25--50~AU, and 0.093 for 50--250~AU.

The work presented in this paper constitutes a first step toward the detection of the population of ``outer'' giant planets around other stars. Such a study, which is complementary to RV searches in terms of orbital separation, is necessary to improve our understanding of the various mechanisms that could generate planets on orbits of tens to hundreds of AU, such as in situ formation triggered by collisions of stars with proto-planetary disks or orbital migration induced by gravitational scattering in multiple planet systems. While the upper limits we have found rule out an important increase in the population of planets at large separations compared to the known population of planets below 3~AU, our sample size and the sensitivities we haved reached are insufficient to tell if the above mechanisms operate at all, and a fortiori which one is dominant. Future searches reaching better sensitivities and targeted at a larger sample of stars will be necessary to answer these questions.

Considerable efforts are currently deployed by major observatories to develop instruments dedicated to the search of giant exoplanets around nearby stars. The Gemini Planet Imager (GPI, Gemini Telescope, \citealp{macintosh_gpi}) and the Spectro-Polarimetric High-contrast Exoplanet Research instrument (SPHERE, Very Large Telescope, \citealp{sphere}) are good examples; they should see their first light in around 2010. These complex instruments will ally an extreme AO system to correct atmospheric wavefront errors to unprecedented levels of accuracy, a calibration system to correct instrumental quasi-static aberrations, a coronagraph to suppress the coherent on-axis stellar light, and differential imaging capabilities enabled by either multi-channel cameras or integral field spectrographs. The expected performance of these instruments, e.g. a contrast better than 17.5~mag at a separation of 0.5\arcsec\ for GPI \citep{macintosh_gpi}, should allow detection of planets of 1~\mjup\ aged less than 100--200~Myr at separations of 5--50~AU, significantly improving on the work presented here. These efforts should uncover the population of outer giant planets, if they exist, or place sufficient constraints on their existence to rule out the mechanisms that could generate them. In less than a decade the James Webb Space Telescope will allow similar studies to be done for relatively nearby M-type primaries, which are too faint for operating the wavefront sensor of extreme adaptive optics systems. Given all of the projects that should unfold in the next few years, the coming decade promises to be extremely exciting for exoplanet science.

\acknowledgments

We are grateful to the referee whose thorough review and excellent suggestions have improved the quality of this paper significantly. The authors would like to thank the Gemini staff for carrying out all the observations. This project was made possible through the support and generous allocation of observing time from the Canadian, US, UK, and Gemini staff time allocation committees. This work was supported in part through grants from the Natural Sciences and Engineering Research Council, Canada, from the Fonds Qu\'eb\'ecois de la Recherche sur la Nature et les Technologies, and from the Facult\'e des \'Etudes Sup\'erieures de l'Universit\'e de Montr\'eal. This research was performed in part under the auspices of the US Department of Energy by the University of California, Lawrence Livermore National Laboratory under contract W-7405-ENG-48, and also supported in part by the National Science Foundation Science and Technology Center for Adaptive Optics, managed by the University of California at Santa Cruz under cooperative agreement AST 98-76783. This research has made use of the SIMBAD database, operated at Centre de Donn\'ees astronomiques de Strasbourg (CDS), Strasbourg, France. This research has made use of the VizieR catalog service \citep{ochsenbein00}, hosted by the CDS.

\end{document}